\documentclass[preprint]{aastex}
\usepackage[dvips]{color}
\newcommand{\ms}{M$_{\odot}$}
\newcommand{\ls}{L$_{\odot}$}
\begin{document}
\title{ Emission Lines from the Gas Disk around TW Hydra and the Origin of the Inner Hole}
\author{U.Gorti\altaffilmark{1,2}}
\author{D.~Hollenbach\altaffilmark{1}}
\author{J.~Najita\altaffilmark{3}}
\author{I.~Pascucci\altaffilmark{4,5} }
\altaffiltext{1} {SETI Institute, Mountain View, CA}
\altaffiltext{2} {NASA Ames Research Center, Moffett Field, CA}
\altaffiltext{3} {NOAO, Tucson, AZ} 
\altaffiltext{4} {Space Science Telescope Institute, Baltimore, MD}
\altaffiltext{5}{Johns Hopkins University, Baltimore, MD}
\begin{abstract}
 We compare  line emission calculated from theoretical disk models  with optical to sub-millimeter wavelength observational data of the gas disk surrounding TW Hya  and infer the spatial distribution of mass in the gas disk.  The model disk that best matches observations has a gas mass ranging from $\sim10^{-4}-10^{-5}$\ms\ for $0.06{\rm AU} <r<3.5$AU and $\sim 0.06$\ms\  for $ 3.5 {\rm AU} <r<200$AU.  We find that the inner dust hole ($r<3.5$AU) in the disk must be depleted of gas by $\sim 1-2$  orders of magnitude compared to the extrapolated surface density distribution of the outer disk. Grain growth alone is therefore not  a viable explanation for the dust hole. CO vibrational emission  arises within $r\sim 0.5$AU from thermal excitation of gas. [OI] 6300\AA\ and 5577\AA\ forbidden lines and OH mid-infrared emission are mainly due to prompt emission following UV photodissociation of OH and water at $r\lesssim0.1$AU and at $r\sim 4$AU. 
[NeII] emission is consistent with an origin in X-ray heated neutral gas at $r\lesssim 10$AU, and may not require the presence of a significant EUV ($h\nu>13.6$eV) flux from TW Hya. H$_2$ pure rotational line emission comes primarily from $r\sim 1-30$AU. [OI]63$\mu$m, HCO$^+$ and CO pure rotational lines all arise from the outer disk at $r\sim30-120$AU. We discuss planet formation and photoevaporation as causes for the decrease in surface density of gas and dust inside 4 AU.  If a planet is present, our results suggest a planet mass $\sim 4-7$M$_J$ situated at $\sim 3$AU. Using our photoevaporation models and the best surface density profile match to observations, we estimate a current photoevaporative mass loss rate of $4\times10^{-9}$\ms\ yr$^{-1}$ and a remaining disk lifetime of  $\sim 5$ million years. 
\end{abstract}
\section{Introduction}
Circumstellar disks are widely believed to go through a transition epoch as they evolve from an initial, gas-rich, classical T Tauri (CTTS) phase to the weak-lined T Tauri (WTTS) debris disk phase (e.g., Strom et al. 1989, Skrutskie et al. 1990, Najita et al. 2007). Transition disks, which typically are a small fraction ($\sim $10-20\%) of all disks, are  characterized by an outer, optically thick dust disk and an inner dust-depleted region or ``hole". However, it is unclear if all disks go through the transition phase (e.g., Currie et al. 2009, Muzerolle et al. 2010). Transition disks are sometimes found to be accreting at rates similar to young disks  suggesting that the dust holes are not devoid of gas, but many objects do not show signs of accretion (Sicilia-Aguilar, Henning \& Hartmann 2010). The duration of the transition epoch is also uncertain (e.g., Muzerolle et al. 2010), although a recent study of the statistics of spectral energy distributions suggests times of $\sim$ a few $10^5$ years (Ercolano \& Clarke 2010). Studying these partially evacuated disks  can potentially provide many insights into how disks evolve and form planets. 

At least three different scenarios have been proposed to explain the morphology of transition disks (see Najita et al. 2007 for a summary). First, the infrared opacity hole may not be a true depletion of material, but the result of dust evolution. Coagulation and grain growth in disks might lead to the formation of larger, ``invisible'' solid objects, thereby emptying the inner regions of small, infrared-emitting dust. In this sequence of events, gas and large solids or planetesimals remain and there is no true mass depletion (e.g., Dullemond \& Dominik 2005).  In a second situation, the hole in the disk is  caused by dynamical interactions with an embedded planet (e.g., Skrutskie et al. 1990, Marsh \& Mahoney 1992, Takeuchi et al. 1996,  Varniere et al. 2006). The Jovian or larger mass planet first creates a gap in the disk and then accretes most of the accretion flow from the outer disk, while a small fraction continues past the planet to accrete onto the star (e.g., Lubow \& D'Angelo 2006). The third explanation for the inner hole is photoevaporation, a process by which high energy radiation from the central star heats the gas to escape speeds and creates gaps (and then quickly central holes) a few AU from the star (e.g., Alexander et al. 2006a, 2006b, Gorti, Dullemond \& Hollenbach 2009, Owen et al. 2010). Determining which of the above three mechanisms dominate in the  creation of the inner opacity hole  may be key to deciphering the role played by transition disks in disk evolution, planet formation, and disk dispersal.   
 
TW Hya is the closest known (at 51 pc, Mamajek 2005) classical T Tauri star with a nearly face-on (e.g., Qi et al. 2006) transition disk that exhibits an inner hole in its (IR-emitting) {\em dust} distribution (Calvet et al. 2002, Ratzka et al. 2007, Hughes et al. 2009).   It is a very well-studied object -- the central star and disk have been investigated from the X-ray to centimeter wavelengths. The X-ray, Far Ultraviolet (FUV; 6eV$<h\nu<$13.6eV) and optical flux from the central star are well known (e.g., Kastner et al. 2002, Herczeg et al. 2002, Stelzer \& Schmitt 2004, Robrade \& Schmidt 2006). The dust\footnote{We refer to solid particles with sizes smaller $\lesssim $1mm as ``dust". Larger solids  are ignored as they have too little surface area and are therefore  not important for gas physics and topics relevant to this paper.} disk has been studied and extensively modeled, with one of the earliest detailed models being that of Calvet et al. (2002, hereafter C02) who inferred the presence of a dust depleted inner cavity of size $\sim 4 $AU from the spectral energy distribution. A very small amount of dust persists in the inner hole, and the observed 10$\mu$m silicate emission feature has been attributed to this component (C02).  Ratzka et al. (2007) also model the dust disk to infer a dust cavity, but with a smaller size, $\sim 0.5-0.8$ AU. The presence of a dust hole, or at least a significantly dust depleted inner cavity, has been confirmed by imaging  and interferometric studies (Eisner et al. 2006, Hughes et al. 2007). The star has a low, but measurable, accretion rate with estimates ranging from $4\times 10^{-10}$ \ms\ yr$^{-1}$ (Muzerolle et al. 2000) to $\gtrsim 10^{-9}$ \ms\ yr$^{-1}$ (Alencar \& Batalha 2002, Eisner et al. 2010), indicating the presence of significant gas in the inner dust hole.

The nature of the transition disk around TW Hya is uncertain.  While a 
dust hole is clearly present and gas must be present in the hole to account for the observed accretion, the amount of gas in this region is not well determined. A radial velocity planet was reported by Setiawan et al. (2008), but located closer to the star ($\sim 0.04$AU) than a planet that could cause the 4 AU dust edge. (This detection has been disputed and the radial velocity variations attributed to stellar surface activity, see Huelamo et al. 2008, Figueira et al. 2010.) TW Hya is known to be a strong far ultraviolet (FUV) and X-ray source and observed to have outward flows traced by [NeII]12.8$\mu$m emission (Herczeg et al. 2007, Pascucci \& Sterzik 2009), which implies that the disk {\em is} presently undergoing photoevaporation. However, photoevaporation theory has traditionally predicted a clean evacuated inner region devoid of dust and gas soon after the formation of a hole, which the observed accretion rate onto the central star contraindicates.  Owen et al. (2011) argue that the inner disk draining epoch, although short ($\lesssim 10^5$ years, the local viscous timescale at $1-4$ AU where the gap opens), may still be observed as an accreting transition disk.

Clearly, spectral line observations of {\em gas} in the TW Hya disk might distinguish between the above scenarios. In recent years, many detections of gas emission lines from TW Hya have become available. The disk around TW Hya was detected in the pure rotational transitions of CO by Zuckerman et al. (1995), van Zadelhoff et al. (2001) and later by Qi et al. (2004, 2006) who also imaged the disk. Herczeg et al. (2002) detected UV emission from H$_2$ from the disk surface near the star.  Emission in the near infrared from vibrationally excited H$_2$ was detected by Bary et al. (2003). Thi et al. (2004) detected sub-millimeter emission from the molecules CO, HCO$^+$, DCO$^+$, CN and HCN.  CO rovibrational emission from the disk has been observed (Rettig et al. 2004, Salyk et al. 2007, Pontoppidan et al. 2008). Resolved line emission of  [NeII]$12.8\mu m$ was reported by Herczeg et al. (2007) and Pascucci \& Sterzik (2009). Bitner et al. (2008), as part of a ground-based survey using TEXES (Richter et al. 2003), failed to detect pure rotational H$_2$ emission from the disk, although Najita et al. (2010) detected emission with the larger beam of {\em Spitzer}.  Najita et al. (2010) present a high-quality {\em Spitzer}  IRS spectrum with numerous emission features of H$_2$, OH, CO$_2$,  HCO$^+$, [NeII], [NeIII] and HI. Thi et al. (2010) recently reported the {\em Herschel} PACS detection of the [OI]63$\mu$m line from TW Hya.  In addition to this rich dataset, strong optical emission has been detected from  [OI] 6300 \AA\  and 5577 \AA\  (Alencar \& Batalha 2002, S.~Edwards, priv. comm.). Pascucci et al. (2011) reanalyze the optical data to also report the detection of emission from [SII] 4069 \AA\ and [MgI 4571 \AA. 

The abundant observational data available for the TW Hya system  make it an excellent target for studying the physical nature of transition disks through theoretical modeling. Gas emission line strengths are very sensitive to the density, temperature and the abundance of the emitting species (e.g., Gorti \& Hollenbach 2008). The emission lines observed largely originate from trace species,  and in order to interpret the data accurately, detailed models of gas chemistry and emission are needed. In this paper, we use our previous disk models (Gorti \& Hollenbach 2004, 2008, Hollenbach \& Gorti 2009,  hereafter GH04, GH08 and HG09 respectively) to  compute  gas emission line luminosities from the disk around TW Hya. In \S 2, we provide a short description of the available data and current understanding in each case. We  briefly describe the modeling procedure and then discuss  disk models that best match observational data in  \S 3.   We consider the implications of our results on the nature of the transition disk around TW Hya (\S 4), discuss the constraints set on disk mass in \S5  and finally present our conclusions in \S 6.

\section{Gas Emission Lines: Observational data}
We summarize below the observed gas line emission at various wavelengths. For clarity of discussion, we do not present the data in chronological sequence of detection, but instead group observations into three wavelength regions, (i) far-infrared and sub-millimeter features, (ii) the mid-infrared (MIR) features, and, (iii)   the optical and near-infared (NIR) features.  Table~\ref{obsdata} lists the line luminosities for  the observed gas emission lines. We do not discuss the dust emission observations, but refer the reader to  C02, Eisner et al. (2006), Ratzka et al. (2007) and references therein. 

\paragraph{Sub-millimeter and far infrared emission:}
Detection of CO 2-1 pure rotational emission from TW Hya was first reported in the seminal paper on gas disk lifetimes by Zuckerman et al. (1995).  Several molecular species (CO, HCN, CN, HCO$^+$, DCO$^+$)  have been detected in the TW Hya disk at millimeter and  submillimeter wavelengths using single-dish telescopes (van Zadelhoff et al. 2001; van Dishoeck et al. 2003; Thi et al. 2004). Wilner et al. (2003) presented the first spatially resolved observations of the disk in the HCO$^+$ J=1-0 line. This was followed by CO rotational line images in the 2-1, 3-2 and 6-5 transitions using the SMA by Qi et al. (2004, 2006).

Thi et al. (2010), using the {\em Herschel} Space Observatory, recently reported the detection of the [OI]63$\mu$m line and set upper limits on the [OI]145$\mu$m and [CII]158$\mu$m lines. In conjunction with $^{13}$CO measurements, they interpret the emission line fluxes as indicative of a low gas/dust mass ratio in the disk ($\sim 2.6-26$) and estimate the disk gas and dust masses as $(0.5-5)\times 10^{-3}$\ms\ and $1.9\times 10^{-4}$\ms\ respectively. 

\paragraph{Mid-infrared emission:}
The [NeII]$12.8\mu$m forbidden line was first reported by Ratzka et al. (2007) from {\em Spitzer} IRS observations along with HI 6-5 and 7-6 emission lines. The [NeII] line was subsequently spectrally resolved using MICHELLE on Gemini N by Herczeg et al.  (2007) and they found  the line profile to be consistent with a single Gaussian centered at $-2\pm 3$ km s$^{-1}$ in the stellocentric frame with an intrinsic FWHM of $\sim~21$ km s$^{-1}$. They concluded that the broad line width could  result from turbulence in a warm disk atmosphere, Keplerian rotation of gas located $\sim 0.1$ AU from the star, or a photoevaporative flow. These observations were followed by high-resolution  spectra using VISIR on Melipal/VLT by Paccucci \& Sterzik (2009) and more recently by Pascucci et al. (2011). They found that the  [NeII] line profile (intrinsic FWHM $\sim 13$km s$^{-1}$)  and the $\sim 4-6$ kms$^{-1}$ blueshift with respect to the star are consistent with theoretical predictions of  a photoevaporative flow driven by stellar extreme-ultraviolet  (EUV) photons (Alexander 2006). The line profile may also be consistent with photoevaporation by X-rays and the production of [NeII] in predominantly neutral, X-ray heated layers (Ercolano \& Owen 2010). 

Najita et al. (2010) recently presented high quality {\em Spitzer } IRS spectra of TW Hya in the $10-20\mu$m region. The spectra  show many emission features including H$_2$ S(1) and S(2) pure rotational lines, many HI lines, [NeII], [NeIII], several transitions of OH and bands of HCO$^{+}$ and CO$_2$. They find no strong lines from  H$_2$O, C$_2$H$_2$, or HCN as seen in other classical T Tauri stars without dust holes (Carr \& Najita 2008, Salyk et al. 2008, Pascucci et al. 2009, Pontoppidan et al. 2010).  They conclude that the [NeII] emission is consistent with origin in the X-ray heated disk surface. The detected OH emission arises from transitions originating in energy levels as high as  23000 K above ground, and they interpret the emission as likely to arise from photodissociation of water in the disk since this process results in OH in very high rotational states (van Dishoeck \& Dalgarno 1984).  The relative strengths of the H I emission are found to be consistent with case B recombination in gas possibly associated with the accretion flow close to the star.

\paragraph{Near-infrared, optical and UV emission:}
Emission from molecular hydrogen in excited vibrational states has been detected from TW Hya. Weintraub et al. (2000) detected emission from H$_2$ in the disk around TW Hya in the 1-0 S(1) line at 2.1218 $\mu$m through high-resolution spectroscopy using CSHELL on NASA's Infrared Telescope Facility.  Herczeg et al.  (2002) detected  H$_2$  UV emission (with FWHM$\sim18$km s$^{-1}$) consistent with   excitation caused by UV-induced fluorescence from a warm surface layer $\sim2-3$AU from the central star. Ingleby et al. (2009) identified a feature in the STIS spectra at 1600 \AA\ which they attribute to electron impact H$_2$ emission and derive a lower limit on the surface density of gas at $\sim$ 1 AU to be $\sim 5\times 10^{-5}$g cm$^{-2}$.

CO rovibrational emission from hot gas, presumably in the inner ($r\lesssim 4$AU) disk, has also been detected. Rettig et al. (2004) first detected CO from the disk, and concluded that the observed emission was optically thin and consistent with a CO gas mass of  $\sim 6\times10^{21}$ g. The low emission temperature of $\sim 430$K was interpreted as due to a clearing of the inner disk out to a radial distance of $\sim0.5$ AU. Salyk et al. (2007) expanded on these results with observations that cover a larger range of rotational energies. They estimate emission radii for the CO gas to be $\sim 0.2-1$ AU, and derive higher temperatures ($\sim 800$K) based on their excitation diagram analysis. Pontopiddan et al. (2008) have recently presented  velocity-resolved spectroastrometric imaging of the  rovibrational lines  (4.7$\mu$m) using the CRIRES high-resolution infrared spectrometer on the Very Large Telescope (VLT). Keplerian disk models are fit to the spectroastrometric position-velocity curves to derive geometrical parameters of the molecular gas and  the CO gas emission from TW Hya is calculated to come from  $r\sim0.1-1$ AU. 

Optical lines from atomic oxygen have been detected from TW Hya. Emission from forbidden lines of [OI] at 6300\AA\  and 5577\AA\ have been observed around many young stars with disks and outflows (e.g., Hartigan et al. 1995). Apart from a high velocity component (HVC, $\Delta v \sim 100-200$km s$^{-1}$) that is attributed to shocks in the outflows, there is also a low velocity component (LVC, $\Delta v \sim 10$km s$^{-1}$) that is believed to originate in the disk or from the base of a slow disk wind.  The origin of this line has remained somewhat of a mystery, as physical conditions in disks are such that reproducing the observed high line luminosities is difficult (e.g., Font et al. 2004, Hollenbach \& Gorti 2009). The higher excitation [OI]5577 \AA\ line often accompanies the LVC, but is never associated with the HVC.
The optical forbidden lines often exhibit small blueshifts (typically $\sim -5 \pm 2 $km s$^{-1}$) (Hartigan et al. 1995). Ercolano \& Owen (2010) explain the LVC emission as arising from an X-ray heated, neutral photoevaporative flow that then gives rise to the observed blueshifts. However, the forbidden line emission in TW Hya is not blueshifted and does not appear to participate in a photoevaporating flow (Pascucci et al. 2011). 
 TW Hya has emission lines (only in the low velocity component) at [OI]6300\AA\ and 5577\AA, with line luminosities estimated at $10^{-5}$ and $1.4\times 10^{-6}$ L$_{\odot}$ respectively (Alencar \& Batalha 2002, Pascucci et al. 2011).  Pascucci et al. (2011) obtain an intrinsic FWHM of the [OI] lines as $\sim 10$km s$^{-1}$  and also report [SII] 4069\AA\  emission, estimated at $3.6\times 10^{-6}$\ls\ and non-detection of the [SII] 6731\AA\ line.  As we show later, very high gas temperatures or densities are needed to excite the [OI] 5577\AA\  line and reproduce the observed ratio of  the 6300/5577 lines ($T_{gas} \gtrsim 30,000$K for $n_e\lesssim 10^6$cm$^{-3}$, or, $T_{gas}\gtrsim 4000$K for $n_e \gtrsim 10^8$ cm$^{-3}$). Thus it is a challenge for the 5577\AA\ emission to arise thermally (however, also see Ercolano \& Owen 2010).   Here we account for the [OI] emission and line ratio  as a result of photodissociation of OH in the disk surface layers (e.g., St\"{o}rzer \& Hollenbach 1998, Acke et al. 2005). 
  
\section{Gas Emission Lines: Modeling }
We use our recently developed theoretical disk models to model the gas emission from the disk around TW Hya. The stellar parameters are reasonably well determined and the dust distribution has been previously modeled, and the focus of this paper is therefore the gas disk modeling.  We vary disk parameters to find the model that best fits the observational data, with  the principal variable being  the surface density distribution.  We find that two disk models with different inner disk masses (for $r<4$AU), but with the same outer disk mass ($r>4$AU), provide reasonable fits to the data. We describe these disk models in what follows,
and discuss in detail the case with the lower inner disk mass as these results are a somewhat better match to the observed line emission. We call this our standard disk model (hereafter referred to as SML, 'L' for low inner disk mass).  We also discuss the results from an alternate model with a higher mass in the inner disk  and refer to that composite model  as SMH. 
 
\subsection{Model description and inputs}
Our theoretical disk models start from a prescribed disk surface density distribution and stellar spectrum and proceed to self-consistently calculate the disk structure from the gas temperature and density distribution.  The gas disk models simultaneously solve for the thermal and chemical properties of the gas and vertical hydrostatic equilibrium. Gas and dust temperatures are computed separately and can deviate significantly in the surface layers where most of the line emission originates.  The gas radiative transfer follows an escape probability formalism.  Dust radiative transfer is done using a modified two-layer approach (e.g., Chiang \& Goldreich 1997, Dullemond, Dominik \& Natta 2001, Rafikov \& DeColle 2006). Our chemistry consists of $\sim 85$ species and $\sim 600$ reactions,   includes photodissociation and photoionization by EUV, FUV and X-rays,
and  is focused towards accuracy in the disk surface layers. Our standard models do not include freeze-out of species on grain surfaces, but we test for the effects of freeze-out where relevant below. We also assume that H$_2$ forms on grain surfaces in our disk models for all grain temperatures below $\sim400$K (Cazaux \& Tielens 2004).  For further details, see GH04, GH08 and HG09.

Stellar parameters and the incident radiation field are key inputs to our theoretical models, and are reasonably well determined for TW Hya. The source is at a distance of 51 pc (Mamajek 2005) and
the disk is almost face-on with a very low inclination angle (7$^{\circ}$ -- outer disk, Qi et al. 2004; 4$^{\circ}$ -- inner disk, Pontoppidan et al. 2008). We adopt $M=0.7$M$_{\odot}$, R=1R$_{\odot}$ and  $T=4200$K  for the stellar mass, radius, and effective temperature (Webb et al. 1999). We construct a composite spectrum for the star by adding observed X-ray and FUV spectra. We use the FUSE/IUE spectrum from the NASA HEASARC archive (Valenti et al. 2003, Herczeg et al. 2002).  The FUV region is split into several energy bins for  accurately computing gas opacity as described in GH04. Herczeg et al. (2002) derive a reconstructed  Lyman $\alpha$ flux $\sim30$\% higher than the observed flux and we use this value for the Lyman $\alpha$ band of our FUV spectrum. The total FUV ($6-13.6$eV) flux is approximately $3\times10^{31}$ erg s$^{-1}$.  The X-ray spectrum is from the XMM-Newton archive (Robrade \& Schmitt 2006) and extends from $0.3-5$keV, and with a total X-ray luminosity of $1.3\times10^{30}$ erg s$^{-1}$.  We extrapolate the observed spectrum to lower and higher energies ($0.1-10$keV) and this results in a total X-ray luminosity of $1.6\times 10^{30}$ erg s$^{-1}$. 
Although the star might also have an EUV component ($13.6-100$eV), perhaps indicated by the observed photoevaporative flow (Alexander 2008, Pascucci \& Sterzik 2009), its strength can only be crudely estimated by indirect arguments (Alexander et al. 2005) and we do not include this uncertain contribution to the high energy flux. We discuss below, where relevant, the effect on line emission fluxes if an EUV flux were to be added. Figure~\ref{spec} shows the composite spectrum used as model input and the vertical dashed lines in the figure indicate the neglect of an EUV radiation field.

Two different dust disk models have been proposed for TW Hya, by Calvet et al. (2002) and Ratzka et al. (2007). 
C02 developed detailed models of the TW Hya  dust spectral energy distribution (SED) to constrain many aspects of the disk structure, and these models have been validated by subsequent observational studies (e.g., Qi et al. 2004, Eisner et al. 2006, Hughes et al. 2007).   The flux deficit at $\lesssim 20\mu$m and a sharp increase or excess at $\sim 20-60\mu$m led them to propose the presence of an inner hole in the disk at $r \lesssim 4$ AU. The deficit is ascribed to a lack of dust in the hole, and the excess arises from direct illumination of the rim of the optically thick dust disk beyond 4 AU. There is also a small mass of dust, $0.5$ lunar masses, in the inner cavity which gives rise to the observed spectral feature at $10\mu$m. Ratzka et al. (2007) situate the inner edge of the outer optically thick dust closer to the star at $\sim 0.5-0.8$ AU. They model  both the SED and the 10$\mu$m visibilities in detail by including sedimentation in the disk, where the dust is reduced in the upper atmosphere due to settling processes.  Our dust models at present do not include settling. We defer gas modeling of the disk using the Ratzka et al. (2007) dust distribution to future work, and keep our dust model  relatively uncomplicated at present. 

  We avail ourselves of the dust models of C02. For simplicity, we adopt many features of this dust model, specifically the dust size and spatial distribution in our standard dust model. We consider their favored model with a maximum grain size $a_{max}$=1cm and disk mass $0.06$\ms, which corresponds to a total dust mass of $7.8\times10^{-4}$\ms\  for their assumed gas/dust ratio.   As we ignore dust particles with sizes larger than 1mm which are not very relevant for gas physics, our model disk ``dust" ($a_{max}=1$mm) mass is reduced accordingly to $2.4\times10^{-4}$\ms (for a grain size distribution $n(a)\propto a^{-3.5}$, where $a$ is the grain size).   Our chemical composition of dust for the outer disk is simpler than the detailed models of C02. We use a mixture of silicates and graphite for the optically thick disk, and verify the assumption by the SED fit from the resulting model dust continuum emission.  We note the disk structure (especially in the regions where gas emission lines originate) is eventually determined by the {\em gas} temperature structure, although the dust distribution is sometimes important for heating and cooling of gas.  We use pyroxene grains for the inner dust hole as suggested by C02 to better fit the continuum SED and the 10$\mu$m feature.  As we show later,  the amount of dust present in this inner region is in any case too low to significantly affect the gas thermal balance or chemical  calculations.  The details of the adopted dust model 
are listed in Table~\ref{calvetdust}.

 Polycyclic aromatic hydrocarbons (PAHs) can efficiently heat gas via the grain photoelectric heating mechanism (Bakes \& Tielens 1994) and gas heating  depends  on the assumed PAH abundance. Disks around low mass stars are believed to be deficient in PAHs by a factor of $\sim 10-100$ (Geers et al. 2006). We typically scale the PAH abundance  with the decrease in  cross-sectional area of the dust  per H atom compared to the interstellar value. We adopt a PAH abundance of $\simeq 10^{-7}$ per H in the ISM (e.g.,Habart et al. 2003). Therefore,  the abundance of PAHs in the optically thick disk ($r\gtrsim 4$AU) around TW Hya is $\simeq 10^{-9}$.  Such a depleted abundance means that X-ray heating usually dominates FUV-induced PAH heating of gas in the outer disk. The  abundance of small grains ($a\lesssim 0.1\mu$m) in the inner disk ($r\lesssim4$AU) is either negligible (C02) or very small (Eisner et al. 2006). It is therefore likely that the abundance of very small dust like the PAHs is low in the inner disk.  Such low PAH abundances do not influence the heating in the inner disk and the PAH abundance in the inner disk  is therefore set to zero.

We do not consider turbulence in the disk beyond 4 AU and all cooling line widths are thermal in the optically thick outer disk.  This assumption is supported at large radii by sub-millimeter observations and modeling of molecular emission from TW Hya by Qi et al. (2006) and Hughes et al. (2010). Hughes et al. (2010) infer a small turbulent linewidth that is  approximately a tenth of the thermal linewidth in the outer disk. The situation may be different in the inner disk, $r\lesssim 1-2$AU.  Salyk et al. (2007) assume higher turbulent linewidths ($v_{turb}\sim0.05 v_K \sim 4$km s$^{-1}$, where the Keplerian velocity $v_K\sim 80$km s$^{-1}$ at 0.1 AU) in their analysis of the rovibrational CO lines from the disk. CO$_{rovib}$ emission arises from hot gas, and the thermal speed of CO for $\sim 2000$K gas is $\sim1$kms$^{-1}$. The expected thermal linewidth is therefore $\sim2$km s$^{-1}$. Observed line widths are typically $\sim 15$kms$^{-1}$ (Salyk et al. 2007), with presumably a large Keplerian contribution for gas located at small radii ($\sim 0.1$AU) in a face-on disk with a small inclination (4$^{\circ}$, Pontopiddan et al. 2008).  We follow Salyk et al. (2007) in assuming $v_{turb}=0.05v_K$ in the inner disk ($r<4$AU) regions. 

We also add viscous heating as a mechanism to heat gas in the inner disk. The dust-depleted inner cavity might still contain dense gas and viscous heating may be an important contributor to heating the dust-poor gas, especially in the shielded interior near the midplane.  We follow the prescription given in Glassgold et al. (2004) (Equation 12 of that paper)  for the viscous heating term. The viscosity parameter $\alpha$ is unknown, but is set in our models by assuming steady state accretion onto the star at the observed rate of $\sim10^{-9}$ \ms yr$^{-1}$ and from the solutions obtained for the density and temperature of the gas.

The surface density distribution of gas in the disk is the primary input to our disk models. We begin  with the dust distribution of C02 which has a rapid increase in surface density over $\sim 3.5-4$ AU. We then vary the gas surface density distribution, $\Sigma(r)$, so that the resulting gas emission lines provide a good match to observational data. The resulting profile for the SML disk is shown in Figure~\ref{sigmar}.  The SML disk has a gas mass $\sim 10^{-5}$ \ms\ within 3.5AU, and 0.06\ms\ for $3.5{\rm AU}<r<200{\rm AU}$. Although the dust surface density drops by $\sim10^3$ inward of $r\sim 4$AU, the gas surface density drops only by $\sim100$ for the SML disk. 
The alternate model SMH has a gas mass  $\sim 10^{-4}$ \ms\ within 3.5AU, and in this case the gas surface density only drops by $\sim 10$. The outer disk surface density distribution is the same for both SML and SMH cases. 
We discuss later in \S3.3 how deviations from the models SML  and SMH worsen the agreement between model results and observational data.

 For descriptive purposes, we divide the disk into three distinct regions, (i) the inner disk ($r\lesssim3.5$AU), (ii) the mid-disk region (3.5AU$\lesssim r \lesssim 20$AU) and the (iii) outer disk ($r\gtrsim 20$AU). Our modeling procedure first constrains the gas in the inner disk region, as this gas provides opacity to stellar radiation, shields gas in the partially exposed rim region at $3.5-4$ AU, and affects the structure of the outer disk.  The inner region of the SML disk is not only depleted in dust but also in gas, by $\gtrsim$ 2 orders of magnitude compared to that expected if the outer optically thick disk was smoothly extrapolated radially inwards. A power law is assumed for the inner disk surface density distribution, $\Sigma(r) \propto r^{-1}$; $0.06$AU$\leq r < 3.5$AU. We place the inner disk edge at $0.06$AU, at the location of the magnetospheric truncation radius (Eisner et al. 2006). As there is continued accretion onto the star, presumably by funnels along magnetic field lines, this is a logical choice for this parameter.  Between $3.5-4$ AU, $\Sigma(r)$ rises sharply and these radii constitute the ``rim" region. The functional form for $\Sigma(r)$ is modeled to fit the appearance of the rim of a photoevaporating disk (Gorti et al. 2009) and is given  by $\log \Sigma(r) = 18 - \exp(11/r_{AU})$ g cm$^{-2}$; $3.5<r_{AU}<4$. Here, $r_{AU}=r/$1AU.  The dust disk turns optically thick over this region. Beyond 4AU, the surface density is assumed to follow a self-similar density profile, as expected from a photoevaporating and viscously evolving disk (e.,g., Hughes et al. 2007, Gorti et al. 2009). For $r>4$AU, $\Sigma(r) = 500 r_{AU}^{-0.7} \exp (-r_{AU}^{1.3}/100)$ g cm$^{-2}$. 

\subsection{Model Results: Disk Structure} We briefly discuss the disk density and temperature structure, the dominant heating and cooling mechanisms at three representative disk radial positions, and then describe the line emission calculated from the disk models. 

The density and temperature distribution as a function of spatial location in the SML disk  is shown in Figure~\ref{nt-contour}.  Gas temperatures in the surface regions where emission lines originate are typically $\sim$  $2000-200$K in the inner and mid-disk regions, $r\lesssim 30$AU.  The near absence of dust in the inner disk and the sudden rise in dust surface density at the rim causes an  increase in gas heating and results in a small vertically extended region at this radius $\sim 4$AU.  However, gas in the inner disk provides opacity that shields the midplane regions at $\sim 4$ AU, and there is  only a slight ``puffing-up" of the inner rim.  (We find that for the upper layers of the outer disk where the gas temperature decouples from the dust temperature, the column density of gas to the star at a given angle (height of the rim) changes by less than 10\% as the mass of the inner disk changes from $10^{-5}$\ms\ to $10^{-4}$\ms.) As seen in Fig.~\ref{nt-contour} (see A$_{\rm V}=1$ surface), the disk structure is overall quite flat, with only modest flaring of the outer disk. Gas at depths where A$_{\rm V}\sim 0.1-1$ to the star, and  at radii $r\gtrsim 50$AU,  only attains temperatures of  $\sim 50-200$K and  the density in these warm layers is low, $n\sim 10^{5-7}$ cm$^{-3}$.

Disk gas temperature and density is shown as a function of height at typical positions in the three different disk regions, inner disk at $r=0.1$AU, mid-disk  at $r=6$AU, and the outer disk at $r=30$AU in Figures~\ref{vslices1},\ref{vslices2} and \ref{vslices3}.  

In the inner disk (Fig.~\ref{vslices1}), heating is mainly by X-rays all through the disk, and in the midplane FUV is significant only via  chemical heating (primarily due to photodissociation of water and ionization of carbon).  However, to attain the high surface temperatures achieved in this region, FUV photons are essential to photodissociate molecules  capable of efficiently cooling the surface gas. The gas temperature increases when H$_2$ photodissociates and molecular (H$_2$O and OH, formed via H$_2$) cooling decreases, and also when CO photodissociates and can no longer cool the gas efficiently. In the inner disk,  the transition from C to CO occurs higher up than the transition from H to H$_2$, as is typical in low dust regions (e.g., Glassgold et al. 1997). 
The inversion in height can be seen in Fig.~\ref{vslices1} where the  H$_2$ and CO photodissociation fronts are marked. 
Rovibrational  emission of CO (CO$_{rovib}$) dominates the cooling with some additional cooling by H$_2$O and OH, until CO photodissociates. At higher $z$, OI6300\AA\  and Lyman $\alpha$ are the main coolants.

Fig.~\ref{vslices2} shows the disk vertical structure in the mid-disk at $r=6$AU. In these regions  where dust is abundant, dust collisions keep  gas at the ambient dust temperature in the optically thick midplane region. Near the optically thin surface layer and near the H$_2$ photodissociation front, heat released from the formation of H$_2$ is significant. Chemical heating due to photoreactions (FUV photodissociation of water) and FUV-induced grain photoelectric heating by PAHs is of significance over a limited column (A$_V\sim 0.1-1$). At the surface, X-rays dominate the heating. Cooling is mainly due to CO and H$_2$ in the molecular layer, and OI6300\AA, [OI]63$\mu$m, [NeII] and [ArII] fine structure lines cool the surface gas.

The outer disk vertical structure at $r=30$AU is shown in Fig.~\ref{vslices3}. Dust is thermally coupled to the gas by collisions in the midplane where gas and dust temperatures are equal. Near the A$_{\rm V}=1$ layer, heating by PAHs, photodissociation reactions and formation and dissociation of H$_2$ are important. X-rays  dominate the heating at vertical column densities of $\lesssim10^{22}$cm$^{-2}$. FUV photoionization of carbon contributes to the heating ($\sim$ 15\%) in the region above the CO photodissociation layer. Main coolants in the outer disk are CO in the molecular region and [OI]63$\mu$m fine structure emission at the surface. 
  
\subsection{Model Results: Line Emission}
We adopt the following procedure to determine the surface density distribution  that is described in \S 3.1.  In order to obtain a gas disk surface density model that best matches observed gas line emission, we first constrain the surface density distribution in the inner disk, i.e. within 4 AU.  This is because gas in the inner disk can shield the outer disk regions, affecting  the density and temperature structure of the disk beyond 4 AU, where the dust disk is optically thick.  
There clearly is some gas in the inner hole, based on the observed accretion onto the central star, and on
observations of CO rovibrational lines and  H$_2$ ultraviolet lines.  We discuss these emission line constraints and the undetected water lines below, and their implications for the gas mass in the inner hole.

\paragraph{Gas Opacity}
Although the dust is optically thin to stellar radiation, the column density of gas may be sufficient to provide significant
optical depth in the inner disk.  In fact, the gas may be optically thick to its own infrared emission.  We use our disk chemical models to obtain chemical abundances for gas in the inner disk and follow Tsuji(1966) in calculating the monochromatic opacity due to each gas species.  At the likely pressures and temperatures of our inner disks, H$^-$ and H$_2^-$ bound-free and free-free, CO and H$_2$O  all can provide
significant opacity.   H$^-$ and H$_2^-$ bound-free and free-free continuum opacity can be significant at wavelengths shorter than about 4 $\mu$m, and produce near IR radiation from the inner edge of the disk at ~$\sim 0.06$ AU.   The ground state CO vibrational band occurs near 4.7 $\mu$m and the blend of CO rovibrational lines provide opacity near 5 $\mu$m.   Beyond about 6 $\mu$m, the blend of pure rotational H$_2$O lines dominate the opacity.   

Later on in this subsection, we discuss CO and H$_2$O lines in some detail. Here, we make more general arguments based on the expected continuum emission from gas in the disk.  We use the opacities from Tsuji (1966) to show that if the surface density of the outer disk is extrapolated  inwards to produce a $\sim 10^{-3}$ M$_\odot$ inner disk (we call this the ``full disk" or the ``undepleted disk"), observable emission would result in the wavelength region $5-15$ $\mu$m due to the CO and H$_2$O emission.    We find, by a similar analysis, that the  H$^-$ and H$_2^-$ bound-free and free-free produce excess emission at shorter wavelengths ($\sim 2-3 \mu$m) where the stellar emission is significant.  We use the specific opacity (per molecule) for each opacity contributor from Tsuji(1966) and calculate the opacity at each radial annulus by multiplying this with the vertical column density as obtained from our model disk chemical calculations. For illustration, we calculate the flux at one wavelength and compare that with the observed emission. We choose a wavelength of 8$\mu$m where the stellar contribution is negligible, there is no contamination from the silicate feature and where there is a significant dip in the continuum emission (see C02).  At each radial annulus, the photosphere (at $8\mu$m) is defined to lie at the height ($z$) above the midplane where the total optical depth at 8$\mu$m becomes of order unity. Near the inner edge of the disk for $0.06{\rm AU} <r \lesssim 0.08 {\rm AU}$, H$^-$ and H$_2^-$ free-free are the main opacity contributors. This opacity drops rapidly with decreasing electron density further out in the disk. At $0.08{\rm AU}\lesssim r\lesssim 0.2$AU, H$_2$O lines dominate the 8$\mu$m opacity.   Having defined the photosphere of the disk from the gas opacities, we estimate the $8\mu$m flux from this surface, where the gas temperature ranges from $\sim 300-2000$K. The disk becomes optically thin to the midplane ($\lambda \sim 8\mu$m) at $\sim 0.2-0.3$AU. We obtain an 8$\mu$m flux of $\sim 1$Jy, a factor of $\sim 4$ higher than observed.  A full gas disk is therefore incompatible with the spectral energy distribution.  The H$_2$O  lines at 8$\mu$m therefore yield an estimate of the upper limit on gas mass as 2.5 $\times 10^{-4}$ M$_\odot$. We discuss later our model results for H$_2$O line emission and compare them to {\em Spitzer} IRS upper limits to the line fluxes to obtain a better limit on the inner disk mass. 

 A similar opacity analysis at a wavelength of 5$\mu$m suggests that CO rovibrational emission in excess of what is observed will be produced by a full gas disk.   We use our model calculations and compare them with observed rovibrational CO line strengths near 5$\mu$m  to estimate the actual mass of the inner disk to be $\sim 10^{-4}- 10^{-5}$ M$_\odot$.

 In what follows, rather than using an opacity argument, we use our disk model results on emergent line fluxes from the disk surface to infer the disk mass.  We describe the emitting regions for each gas emission line diagnostic and also discuss where applicable the constraints set in each case. The outer disk gas is attenuated by the gas in the inner disk as described earlier. We calculate the outer disk structure using  the surface density distribution of the inner disk in the SML model and use the {\em same} results for the outer disk component of the SMH model as well, as the attenuation columns of the gas provided at 4 AU are similar in both these cases (within $\sim$10\%). As our disk models are computationally intensive, we do not calculate disk models for the outer disk in the SMH case separately. In the rest of the paper, the emission in the  mid and outer disk regions refer to that calculated using gas optical depths and absorption columns for the $10^{-5}$\ms\ inner disk mass case (SML).   Figure~\ref{lines} shows the radial extent of each of the gas emission lines discussed below. The cumulative line luminosity, $L(r)$, integrated up to the radius $r$ is shown for each emission line and is normalized to the total calculated value (from inner, mid and outer disk regions). Also marked are the 10\% and 90\% luminosity levels. We quote the corresponding radii below for each emission line and this defines the radial extent of the emission.

\paragraph{CO rovibrational emission}

CO$_{rovib}$ emission typically arises from a height in the inner disk where the CO is just beginning to photodissociate. Gas temperatures in the disk generally increase with height and  gas temperatures in the emitting layers are high ($T_{gas} \gtrsim 500$K).   In dust-poor gas, the CO and H$_2$ photodissociation layers are inverted, as first reported by Glassgold et al. (2004), that is, the carbon becomes CO at larger $z$ than the
H becomes H$_2$.  We similarly find that the H$_2$ is photodissociated to deeper layers and that CO is co-existent with large columns of atomic hydrogen, creating conditions favorable for exciting CO$_{rovib}$ emission (Glassgold et al. 2004). The primary formation route to CO is the oxygenation of atomic carbon (C+OH). In the layers shielded from FUV photodissociation CO is destroyed by collisions with helium ions. 

CO vibrational lines are an important constraint for setting the gas mass in the inner disk. A high disk mass results in  high total CO vibrational emission if the lines are optically thick, as the luminosity in these lines increases with temperature which increases (albeit weakly) with increasing gas mass. The total luminosity of the observed lines (select P and R branch transitions)  is constrained by observations  to be $\lesssim 10^{-5}$\ls\ (Salyk et al. 2007). We find that inner disk masses of $10^{-4}, 10^{-5}$ and $10^{-6}$ \ms\ all give total CO luminosities within a factor of $\sim 2$ of the observations and are all hence viable choices.  The CO emission is found to be mainly a surface effect dominated by hot gas at the surface of the disk, resulting in approximately the same total CO luminosity for all three disk masses.  
To distinguish between these cases, the line luminosity of each of the CO rovibrational lines needs to be compared with data.   We estimate the CO rovibrational line luminosities of the $v=1-0$ P(J) series as follows. We adopt the analysis of Scoville et al. (1980) and perform an approximate  non-LTE calculation for the populations of the CO rovibrational levels. We only consider the $v=0$ and $v=1$ states as higher vibrational states are unlikely to be significantly occupied at typical disk temperatures. We assume that collisions result in a change in the vibrational state, and vibrational level populations are determined from detailed balance assuming an ``effective'' Einstein A-value from $v=0-1$ as described in Najita et al. (1996). We then assume that the rotational levels within each $v$ state are thermally populated (Scoville et al. 1980). Collisional rates with $e^-$, H and H$_2$ are taken from Najita et al. (1996).  Oscillator strengths and transition frequencies are from the compilations of Kirby-Docken \& Liu (1978).  The optical depth for each transition is calculated as detailed in Hollenbach \& McKee (1979) from the column density of CO molecules in a given rovibrational state,  and an escape probability formalism is used for line transfer. 

  Salyk et al. (2007) detect 15 transitions covering rovibrational energies $E_u/k$ from 3000K-6000K, where $E_u$ is the energy of the excited upper level. We also compare our results with earlier CO rovibrational detections by Rettig et al. (2004) in complementary transitions. There was a typographical error in the Rettig et al. (2004) table of line fluxes, and the actual fluxes are a factor of 10 below what has been reported. We also do not include the P(21-25) transitions of that paper, as these data are prone to errors due to photospheric and telluric corrections (Sean Brittain, private communication).  
We compare our model results with the data in Figure~\ref{covib}.   We find that a gas depletion from the full disk of $0.1-0.01$ corresponding to a gas mass of $10^{-4}-10^{-5}$M$_{\odot}$ matches the observed line emission fairly well, with the lower mass inner disk model (SML) being a better fit to the CO data.  A factor of 10 lower in mass results in CO emission from  warmer gas that produces excess emission at high J levels. 
All of the CO gas emission comes from $r\lesssim 0.5$AU (due to rapidly decreasing gas temperatures beyond), in agreement with both the analysis of Salyk et al. (2007) and the spectroastrometric imaging study by Pontopiddan et al. (2008).  Our gas mass estimate for the SML disk is  consistent with the surface density estimate of Salyk et al.(2007) of 1 g cm$^{-2}$ at $r\sim1$ AU where they used a slab model, although in our models the emission is mainly from hot gas in the disk surface layers.  Salyk et al. (2007) also assume a  H$_2$/CO conversion ratio of 5000 as observed  towards a massive protostar, and caution that their derived mass estimate is likely be a lower limit. In our disk models, the hydrogen is predominantly {\em  atomic} in the regions where CO$_{rovib}$ emission originates (also see Glassgold et al. 2004).   We explicitly calculate  disk chemical abundances and find that the H/CO ratio in the hot inner disk regions where CO$_{vib}$ emission arises is  $\sim$7000.    The gas/dust mass ratio in the inner disk is $\sim 500-5000$ for the dust mass of the Calvet et al. (2002) disk, indicating a relative depletion of the dust with respect to the gas at $r\lesssim 4$AU. 
 
\paragraph{H$_2$O rotational emission}
 Our gas opacity analysis above suggests that a full undepleted disk would produce excess H$_2$O rotational emission in lines whose wavelengths range from 6 to 15 $\mu$m compared with observations. As we specifically calculate water abundances and the disk density and temperature structure, we can compare the model disk H$_2$O rotational line fluxes to the corresponding upper limits from {\em Spitzer} IRS in the $10-19\mu$m wavelength region (Najita et al. 2010). No water emission has been detected and line luminosities are therefore expected to be lower than $\sim 10^{-7}$L$_{\odot}$. We find that for the full undepleted inner disk with a mass of $10^{-3}$\ms, many rotational lines in the $10-19\mu$m region are above the detection limit. Some of the strongest transitions are $7_{61}-6_{16}, 7_{62}-6_{15}, 8_{54}-7_{07}, 8_{53}-7_{26}$ and $8_{44}-7_{17}$, all with luminosities in the range of $1-3\times 10^{-6}$L$_{\odot}$, compared with the upper limits set by {\em Spitzer} observations of $\sim 10^{-7}$L$_{\odot}$. 
From the water rotational line emission constraint alone, we can set a limit on the inner disk mass as $\sim 10^{-4}$\ms, at which  the water emission is barely detectable. This limit is consistent with the upper limit set by CO  emission in the preceding discussion, where we fix the inner disk mass to be in the range $10^{-4}-10^{-5}$\ms\ based on  CO rovibrational emission.

\paragraph{H$_2$ fluorescent emission}
We do not model  H$_2$ UV fluorescence  but draw on the results of Herczeg et al. (2004) who conduct a very detailed analysis and model the observed emission. Our SML disk corresponds to their thick disk model, and most of the emission in this case arises from the illuminated edge of the disk at $r=0.06$AU. Herczeg et al. (2004) estimate a hot ($T_{gas}\sim2500$K), H$_2$ mass of $\sim 10^{19}$g from their analysis. We calculate the mass of H$_2$  gas at these temperatures in our inner disk to be $\sim 3\times 10^{19}$g, in reasonable agreement with the Herczeg et al. (2004) results.  All of this gas lies very close to the inner edge at $\sim 0.06$AU.  At this radius the Keplerian speed of the gas from the almost face-on disk ($i=4^{\circ}$) is about $\sim7$km s$^{-1}$, the thermal speed is $\sim 4$km s$^{-1}$, bringing the expected FWHM to $\sim16$km s$^{-1}$ very close to the observed FWHM of $18$km s$^{-1}$ (Herczeg et al. 2002).   
We note that the inner disk structure near the magnetospheric truncation radius is likely to be quite complicated compared to our simple power law assumption of the surface density, and our comparison with the Herczeg et al.  model is only a consistency check.

\paragraph{OH lines}
16 rotational transitions of OH, mainly in the ground vibrational state, were reported by Najita et al. (2010) from their {\em Spitzer} observations. The data indicate high excitation states (J$_{upper}\sim$ 15-30), with equivalent energies $E_u/k$ up to 23,000K above ground.  OH mid-infrared {\em thermal} emission from our model is mainly from the hot, dense ($n\sim 10^{12}-10^{13} {\rm cm}^{-3}, T_{gas}\sim 700-1000$K) gas very close to the central star at $r\lesssim0.1$AU. These lines are, however,  weak in the $10-19\mu$m region and below the sensitivity of the {\em Spitzer} IRS instrument.   OH thermal emission is also produced from gas in the mid-disk and outer disk regions, but this gas is typically at  $\sim 300$K and incapable of reproducing the observed high excitation lines. As suggested by Najita et al. (2010) for TW Hya,  UV photodissociation of water is a possible source of the OH emission (see also Tappe et al. 2008 who suggest this mechanism for HH214). 

From our disk chemical network solution, we can estimate the total luminosity in the OH lines produced by the photodissociation of water in the disk. We calculate the total OH line luminosity as follows. For simplicity,  in this discussion we assume that Lyman $\alpha$ photons ($E=10.2$eV) are mainly responsible for the photodissociation of water molecules  -- a reasonable assumption because  Lyman $\alpha$ contributes to  a significant fraction of the total FUV luminosity (e.g., Herczeg et al. 2002, Bergin et al. 2003).  Harich et al. (2000) studied the photodissociation of water by Lyman $\alpha$ photons and the photodissociation energy is determined to be 5.1eV. The excess energy for a Lyman $\alpha$ photodissociation is therefore 5.1eV.  About 66\% of the photodissociations of water result in OH and H,  while the remaining 34\% result in  O and H atoms.  Harich et al (2000) further find that most of the OH products are extremely rotationally excited with a peak at a rotational level $J=45$ ($E_u/k \sim$ 45,000K) and that almost 75\% of the available energy goes into pure rotational excitation. This translates to an energy of $\sim 4$eV per photodissociation (of H$_2$O to OH) that is then eventually radiated away in a rotational cascade from the high J states (also see Mordaunt et al. 1984, van Harrevelt \& van Hemert 2000).  We calculate the energy in the rotational cascade in our model disk by counting the total number of water photodissociations by photons with energy equal to or greater than  Lyman $\alpha$  that lead to OH and multiplying this by the energy available for excitation (4eV). If $\dot{N}_{PD}(H_2O)$ is the rate of the total number of photodissociations of water in the disk by Lyman $\alpha$ (and more energetic FUV) photons, then the  luminosity in the OH cascade $L_{OH,PD}$ is the product $0.66 \times \dot{N}_{PD}(H_2O) \times 4$eV. Equating this to the observed OH luminosity then requires that there are $\sim 10^{40}$ photodissociations per second, or that only $\sim 1$\% of the stellar Lyman $\alpha$ luminosity of $10^{42}$ photons s$^{-1}$ be intercepted by water in the disk (also see Bethell \& Bergin 2009). If the solid angle subtended by the disk is $\sim 0.1$,  the desired Lyman $\alpha$ flux absorption  by water
is feasible  if water abundances are high enough (at the 10\% level) to compete  with the other dominant sources of opacity such as dust, photoionization ionization of Mg, Fe, and Si, and photodissociation of O$_2$ and OH. Therefore, photodissociation  of water can plausibly account for the observed line luminosities. 

 In the above approximate analytic calculation, we simply used the Lyman alpha flux. However, in the model  we calculate the OH luminosity due to water photodissociation using photodissociation rates obtained at each spatial grid cell. We use the full FUV spectrum, including Lyman alpha, but only count those photodissociations where the FUV photons are energetic enough to
lead to OH in highly excited rotational states as described above. The OH luminosity in the SML disk is $4\times10^{-6}$\ls,  lower than the 
total luminosity in the observed lines (upper J levels ranging from $\sim30$ to $\sim15$), $6\times10^{-6}$L$_{\odot}$. We note that the cascade must also produce lines that are not observed in the {\em Spitzer} IRS band. The energy of a rotational state $\propto J(J+1)$ and the cascade originates at $J\sim45$. The unobserved transitions from $J\sim45$ to $J\sim 30$ and $J\sim15$ to $J=0$ therefore have a total energy approximately 1.5 times that in the transitions  that are observed.  The unseen luminosity must account for another $\sim 9\times10^{-6}$L$_{\odot}$, bringing the ``observed" value to $1.5\times10^{-5}$L$_{\odot}$.  Our SML disk results for OH produced by H$_2$O photodissociation are  discrepant by a factor of $\sim 4$. Of this emission, 30\% arises from photodissociation of water in the inner disk and the rest comes from the extended middle and outer disk regions to radii $r\lesssim 40$AU (Fig.~\ref{lines}). In the inner disk, the OH is from the edge of the disk near the star where the number densities are $\sim 10^{12}-10^{13}$  cm$^{-3}$ and gas temperatures are $\sim 500-1000$K.  The main formation route to water is H$_2$ + OH and it is destroyed mainly by FUV photodissociation. The water abundance in these regions is however low, with  typical vertical column densities $\lesssim 10^{15}-10^{16}$ cm$^{-2}$ in the photodissociation layer. In the mid-disk and outer disk regions, OH emission is dominated by the rim at 4 AU and beyond the rim is primarily produced from  regions where the column density to the star $\sim 1-5\times 10^{22}$ cm$^{-2}$, densities are $10^7-10^{10}$ cm$^{-3}$ and $T_{gas}\sim200-600$K. In the mid-disk regions OH+H$_2$ is the dominant formation route to water, whereas in the cooler outer disk regions ($r\gtrsim 15$AU), H$_3$O$^+$ recombination leads to the formation of water molecules. Water abundances range from $X(H_2O) \sim 10^{-6}$ to $10^{-9}$ in these photodissociation layers, and typical water columns to the star are $\sim 10^{16}$ cm$^{-2}$.   The latter is consistent with our earlier analysis, which suggested that if the water opacity were $\sim 10\%$ of the total opacity at FUV wavelengths, we would obtain OH luminosities close to that observed.   A column density of $\sim 10^{17}$ cm$^{-2}$ of water provides optical depth of unity at the FUV wavelengths needed to photodissociate water. 

If the OH is mainly prompt emission following photodissociation of water, the SML disk appears to produce less water by a factor of $\sim 4$ than required by the observations. We consider a few mechanisms by which  the water production rate may be enhanced to match observations. Raising the gas temperature and hence increasing the water formation rate by the endothermic H$_2$+OH reaction is one possibility and this could be achieved by increasing the unknown PAH abundance and FUV grain photoelectric heating in the disk.   However, as H$_2$ pure rotational emission originates at the same spatial location where there is warm molecular gas, any increase in temperature overproduces the H$_2$ S(1) line flux.  We discuss the H$_2$ S(1) and S(2) line fluxes below.

Another possibility is that the SMH disk is the better solution for the inner disk mass. This disk is a reasonable match to the CO lines (Fig.~\ref{covib}) and does not violate other observational constraints such as the lack of mid-IR water emission.  Higher densities in the inner disk result in higher water abundances and the prompt OH emission increases by a factor of $\sim 4$ in the inner disk, bringing the total OH luminosity to $\sim 7.8\times 10^{-6}$ L$_{\odot}$, and only a factor of two below the observed value (see Table~\ref{idtable}). This model slightly overestimates the OI6300\AA\  and H$_2$ S(2) line luminosities, but may nevertheless be a feasible model for the TWHya disk. The SMH disk also has the advantage that it better fits the [OI]6300\AA\ linewidth and provides a more typical value for the turbulence parameter $\alpha$ in the inner disk (see Table~\ref{idtable}). 

Water formation on dust grains is another mechanism that could increase water in the disk. Our disk models do not include freeze-out and desorption processes to properly consider grain-surface reactions. We estimate an upper limit to the effect of water formation on dust in a simple manner as follows. We assume that oxygen atoms stick on grains, are instantly hydrogenated and released into the gas phase as water,  at  a rate equal to the collision rate of O with grains. We applied dust formation of gas phase water only in regions where 
 the dust is cold enough to prevent thermal evaporation of O or OH before  H$_2$O is formed on the grain surface, $T_{gr}>100$K. We use this relatively high grain temperature for thermal desorption so that we obtain an upper limit on the production of water on grains. We also require that  $A_V$ to the star is less than $\sim 3$ so that photodesorption can clear the
water ice from the grain surface (Hollenbach et al 2009).   Glassgold et al. (2009) considered a similar mechanism to increase water production in disks but found that it was not significant in increasing column densities of warm, observable water. However, the gas temperature is not relevant for our calculation as the OH lines are a result of  photodissociation and not thermal emission.  Water formation of grains in the SML disk (with $10^{-5}$\ms\ within 4AU) leads to a  higher OH luminosity,  $7\times10^{-6}$\ls, but still a factor of $\sim 2$ below what is observed. We emphasize that this  $7\times10^{-6}$\ls\ is an upper limit, given the optimistic assumptions on water formation on grains. 

Radial transport of solids or water-bearing ices in the disk, a process not considered by us, has been suggested as a possible agent causing local enhancements and depletion of water abundances in the disk (Ciesla \& Cuzzi 2006). Water ice from the midplane may also be transported to greater heights where it is subject to thermal desorption or photodesorption, leading to higher water abundances in the disk surface where it can be photodissociated.  These effects, while likely to operate in disks, require time-dependent models that consider radial and vertical transport and are beyond the scope of this paper. 
However, one simple calculation is illuminating.  If water is transported as water ice radially inwards to inner regions where it is photodissociated once, the mass influx needed is $10^{40}$ H$_2$O molecules per second. Assuming a H/H$_2$O ratio of 2000 (if every O atom is in H$_2$O ice) this translates to $2\times 10^{43}$ H atoms per second or $5\times10^{-7}$\ms\ yr$^{-1}$. Since TW Hya has an age of 10Myr,  and with a constant influx, this requires an improbable initial mass reservoir of 5\ms\ in the outer disk and also implies the current disk will only last $10^5$ years. This is  unlikely.  On the other hand, if turbulent mixing were to act vertically, then this has the advantage that the water can be reformed at the midplane and brought to the surface to be photodissociated repeatedly. For our adopted gas mass of 0.06\ms, we therefore require that this mass be brought to the surface of the disk where it is photodissociated every $10^5$  years. At a typical radius, $r=10$AU, the distance from midplane to the $A_V=1$ surface layer where water is photodissociated is $\sim 1$AU. Thus, the average vertical transport speed is only 1AU$/10^5$yrs$=5$ cm s$^{-1}$. This suggests that vertical mixing may supply the needed water. 

We have discovered a new mechanism for the rotational excitation of OH that may be the most promising, but difficult to quantify precisely at this time. Here, we do not need a greater production of gas phase water. The process is initiated by the photodissociation of OH, which leads to atomic oxygen in an excited electronic state (O$^1$D) approximately 50\% of the time (van Dishoeck \& Dalgarno 1984). The $^1$D state of oxygen, due to its empty 2p orbital,  is  more electrophilic than the triplet ground state $^{3}P$ and is highly reactive to readily undergo bond-forming addition reactions.  O($^1$D) + H$_2$  re-forms OH very efficiently (with a rate coefficient $\gamma_O = 3.0\times10^{-10}$ cm$^3$ s$^{-1}$; NIST Chemical Kinetics database) leading to an effective reduction in the overall destruction of  OH.  This route has no thermal barrier as in the reaction of  the ground state O($^3$P) atoms with H$_2$ and  is especially important at high gas densities when the rate of formation of OH by this route is faster than the radiative decay of the O$^1$D  atom to the ground state (with $A\sim 8\times 10^{-3}$s$^{-1})$.  The fraction $f_O$ of O$^1$D that reacts with H$_2$ before radiatively decaying to the ground state is given by $f_O=1/(1+(n_{crit}/n(H_2)))$, where the critical density $n_{crit} \sim \gamma_O/A = 3\times 10^7$cm$^{-3}$.  An important consequence of the  O($^1$D) + H$_2$ reaction may be that this reaction produces OH in a rotationally excited state (with an efficiency $\epsilon_{OH} \sim 0.2$ at J$\sim$20-30 states, Lin \& Guo 2008) which could then cascade to lower J to produce the observed MIR emission by {\em Spitzer}.  In regions with $n(H_2)>n_{crit}$,
photodissociation of OH itself (rather than H$_2$O as considered earlier) may result in the re-formation of rotationally excited OH. We estimate the contribution of this mechanism to the OH line emission as follows. 

$\dot{N}_{OH}$ photodissociations per second of OH will produce  $\sim 0.5 \epsilon_{OH} f_O \dot{N}_{OH}$ OH molecules in the J$\sim20-30$ state per second.    For our SML disk, we estimate that this mechanism could result in $\sim 5\times 10^{-6}$\ls\ (nearly all from the inner $r\lesssim1$AU region), which recovers most of the observed OH emission which the model was deficient in when we considered the photodissociation of water alone. We will explore this mechanism  for producing OH emission in greater detail in a future study where we will treat O$^1$D as a separate species for better accuracy, and treat the efficiencies of $\epsilon_{OH}(v,J)$ individually.

We conclude with a footnote to the above discussion. Mandell et al. (2008) recently reported the discovery of OH from two HAeBe stars and concluded that the emission was a result of OH fluorescence. The rotational energy diagram of the transitions observed (in this case from the  $v=1$ state) indicate a single rotational temperature which leads the authors to conclude that it is unlikely to be the result of "prompt" emission, i.e., OH produced by the photodissociation of water. Their models of fluorescent emission from  OH match the observed data from HAeBe stars quite well. In their models, OH is excited to the v=1 state through rovibrational transitions pumped by near infrared photons and electronic transitions pumped by UV photons. However, the {\em Spitzer} data  for TW Hya indicate mainly a downward rotational cascade from the high J states of $v=0$ state (Najita et al. 2010). Therefore, in TW Hya the observed  OH is more likely  to be prompt emission.

\paragraph{H$_2$  rotational lines}
Pure rotational S(2) and S(1)  line emission from H$_2$  was reported by Najita et al. (2010) at $4.8\times 10^{-7}$ and $10^{-6}$ L$_{\odot}$ respectively. Most of the H$_2$ emission in our models comes from within radii of $20-30$AU, with $\sim 35$\% of the  H$_2$ S(2) emission and 10\% of the H$_2$ S(1) emission from the inner SML disk (see Fig.~\ref{lines}). In the mid-disk region, H$_2$ emission originates  in the superheated dust later at $A_V\sim0.3-0.1$  and where the gas temperatures are $100-500$K.
H$_2$ formation in the mid and outer disk ($r>4$AU) is mainly on grains and it is destroyed by FUV photodissociation.  We obtain H$_2$ S(2) and S(1) line luminosities  to be $3.4\times 10^{-7}$ and $1.2\times 10^{-6}$L$_{\odot}$ respectively. These are in reasonable agreement with the {\em Spitzer }IRS measurements. Bitner et al. (2008), in their ground-based survey for H$_2$ using TEXES, did not detect H$_2$ S(1) emission and obtained an upper limit of $5.0\times10^{-7}$L$_{\odot}$. This suggests that half of the S(1) emission observed by {\em Spitzer} IRS must originate from a more extended region outside the narrower slitwidth of TEXES, i.e. at $r\gtrsim 20$AU. Although the H$_2$ S(1) emitting region is quite extended (Fig.~\ref{lines}), half the H$_2$ S(1) emission in the SML disk in fact originates at $r\lesssim14$AU. The S(1) flux within 20 AU is $8\times10^{-7}$L$_{\odot}$ and the radial extent of emission is in moderate agreement with the TEXES and {\em Spitzer} data. The line luminosity in the H$_2$ S(0) transition is predicted from the SML disk to be $3\times10^{-7}$\ls.
   
\paragraph{[OI]6300\AA\ and 5577\AA\ emission} We show below that the [OI]6300\AA\  line from TW Hya originates from the disk, mainly from the photodissociation of OH (also see St\"{o}rzer \& Hollenbach 1998, Acke  et al. 2005) with a small contribution from thermal emission. The line luminosities for the [OI]6300\AA\ and [OI]5577\AA\  lines from  TW Hya are $ 10^{-5}$L$_{\odot}$ and  
$1.4\times 10^{-6}$L$_{\odot}$ respectively (Alencar \& Batalha 2002, Pascucci et al. 2011, S.Edwards, private communication). We first show that the observed [OI]6300\AA/[OI]5577\AA\  line ratio of $\sim 7$ makes it unlikely that the origin is thermal. 

The oxygen atom has a triplet ground state ($^3$P) and singlet D and S states which are at $E_u/k \sim 22,850$K and $ 48,660$K respectively. The OI 6300\AA\ transition from the $^1$D to the ground state and the OI 5577\AA\ transition from the $^1$S state to the  $^1$D state therefore require high gas temperatures for collisional excitation to these high energy levels and subsequent radiative
 de-excitation. If the gas is collision dominated (in LTE), then the ratio of the two lines can be easily obtained from the ratio of $n_u A \Delta E$, where $n_u$ is the upper level population, $A$ is the Einstein A-value and $\Delta E$ is the energy of the photon emitted. In LTE, $n(^1S) \sim (1/5)  n(^1D) e^{\Delta E/kT}$, where $(1/5)$ is the ratio of the statistical weights and $\Delta E/k \sim 26,000$K  is the energy difference between the  two levels. Therefore
\begin{equation}
{ {L_{6300}}\over{L_{5577}}} =  { {n(^1D) A_{6300} \Delta E_{6300}}\over{ n(^1S) A_{5577} \Delta E_{5577} }} \sim 5.4 \times 10^{-3} e^{26000/T}
\end{equation}
where $A_{6300}\sim 6\times 10^{-3}$ s$^{-1}$ and $A_{5577}\sim 1.3$ s$^{-1}$. To obtain the observed line ratio of 7,  gas temperatures therefore need to be $\sim 3600$K, which is possible with X-ray heating of the disk surface. However, the gas densities have to be high enough to produce LTE conditions.  Collision partners can either be electrons or neutral hydrogen atoms and typical collision rate coefficients  are $\gamma_e \sim 5\times 10^{-9} (T/10^4)^{0.6}$ cm$^{3}$ s$^{-1}$ for electrons for both the transitions. The collisional rate for O($^1$D) with H atoms is $\gamma_H \sim 10^{-12}$  cm$^{3}$ s$^{-1}$ (Krems, Jamieson \& Dalgarno  2006). $\gamma_H$ for O($^1$S) is not known. However,
 the rates  for the relaxation of the O($^1$S) state are expected to be smaller than those for the relaxation of the $^1$D state, because the spin-orbit couplings in  electronically excited atoms are generally weaker (R. Krems, priv. comm.). We conservatively assume $\gamma_H \sim 10^{-12}$  cm$^{3}$ s$^{-1}$ for O($^1$S) as well.   Therefore, electrons dominate collisions if $x_e>4\times10^{-4}$. For levels to achieve LTE,   electron densities $n_e > A/\gamma_e \sim 10^8$ cm$^{-3}$  are needed for the 5577\AA\ line ($n_e > 10^6$ for the 6300\AA\ line). Collisions with H atoms require neutral densities about 2500 times these values. Even for gas with high electron fractions $x_e\sim 0.1$, gas densities therefore need to be high, $n_H \gtrsim 10^9$cm$^{-3}$. These conditions are likely to be attained only  in the inner disk, where gas at the surface may be dense, be partly ionized with sufficient neutral oxygen, and at $T\gtrsim 3600$K. 

For gas at lower densities  a balance between radiative de-excitation and collisional excitation and de-excitation of the different levels determine the level populations. We calculate the line ratios as a function of electron density and gas temperature and Figure~\ref{o1ratio} shows the contours for the observed value of 7 and other values about this number. From the contour plot, one can see that even for relatively high gas temperatures of $\sim 8000$K  the required electron densities are $\sim 10^{7}$cm$^{-3}$ implying gas densities of $\sim 10^9$cm$^{-3}$ for an electron fraction of $x_e\sim 0.01$, which is typical.  Such conditions are unlikely, except perhaps in the innermost regions, $r\lesssim 0.1$AU. We also note that soft X-rays that can heat the gas to such high temperatures typically have very small absorption columns $\sim 10^{19}$cm$^{-3}$, and very little mass is expected in gas at these high temperatures and densities. Such low masses may not be able to reproduce the observed high optical OI line luminosities.

For typical densities in the hot disk surface, $n\sim 10^{5-7}$cm$^{-3}$, and electron fractions $x_e \sim 0.01-10^{-3}$ the electron densities are $n_e\lesssim 10^5$cm$^{-3}$, and the observed line ratio requires $T\gtrsim 30000$K.  We consider this extremely unlikely as it is very difficult to heat disk gas to these high temperatures and moreover retain a significant amount of neutral oxygen. The recent calculations of Ercolano \& Owen (2010) also show that the 5577\AA \ line is appreciable relative to the 6300\AA\ line only 
when gas densities are high, as in their primordial disk models where the OI emission is from the inner $r\lesssim 1$AU region. In their models, the OI emission is thermal in origin and they quote $ L_{6300}/L_{5577} $  $\sim13-50$ for the primordial disk models and $\sim 300-700$ for the transition disk models, supported by the above analysis.  Only one of their many models, a primordial disk with $L_X\sim 2\times10^{30}$erg s$^{-1}$ has a low line ratio $\sim 13$,  where the emission presumably originates from gas with high electron density.

In  the SML disk, thermal emission contributes to about 20\% of the observed  [OI]6300\AA\ line and 10\% of the 5577\AA\ line. In the inner disk, the luminosities are $1.2\times 10^{-6}$L$_{\odot}$ and $10^{-7}$L$_{\odot}$ for the thermal component of the 6300\AA\ and 5577\AA\ lines respectively, 
and the emission comes from regions where $n_e \sim 10^{5-8}$cm$^{-3}$ $T\sim 4000-8500$K at the disk surface. In the SMH disk, the calculated thermal [OI]6300\AA\ line luminosity is higher, $2\times 10^{-6}$L$_{\odot}$ (Table~\ref{idtable}). There is also thermal [OI]6300\AA\ emission from the mid and outer disk surface that is heated by X-rays to $T\gtrsim 5000$K, the luminosity here is $\sim 1.1\times 10^{-6}$L$_{\odot}$. There is negligible thermal [OI] 5577\AA\ emission from the mid and outer disk regions. 

We propose here that most of the OI forbidden line emission from TW Hya is the result of photodissociation of OH in the disk. 
The OH photodissociation layer in the SML disk is almost spatially co-existent with the H$_2$O photodissociation layer, and this generally is located near the H/H$_2$ transition in the disk. OH photodissociates about 50\% of the time to produce an oxygen atom in the $^1$D state, and occasionally (5\% of the time) in the higher $^1$S state (e.g., Festou \& Feldman 1981, van Dishoeck \& Dalgarno 1983). Subsequent electronic de-excitation to the ground state produces the oxygen doublet at 6300\AA\ and 6363\AA\ ($^1$D$-^{3}$P) and the $^1$S$-^{1}$D transition produces the  5577\AA\ emission. The atoms in the O$^1$D state produce the 6300\AA+6363\AA \ doublet in the branching ratio of $3$ (Storey \& Zeippen 2000).  The ratio of these line strengths is the same as the branching ratio because both lines originate from the same upper state. These calculations  are  supported by comet observations of the photodissociation products due to solar Lyman $\alpha$ (Morgenthaler et al. 2007). We again count the number of photodissociations of OH in the model disk as we did for water, and multiply this by the energy of the transition and the theoretically calculated branching ratio. 

Our inclusion of the  O($^1$D) + H$_2$ $\rightarrow$ OH reaction leads to a lower effective photodissociation rate of OH as discussed earlier.  If this reaction were not considered, then prompt emission produces a 6300\AA/5577\AA\ line ratio $\sim$ 7. However, in regions where the densities are high, such as in the inner disk, the line ratio is expected to be lower as the reactive nature of the O$^1$D atom will result in a suppression of 6300\AA\ emission. Although the $^1S$ state of oxygen is also more electrophilic than the triplet ground state, the unpaired electrons likely make the O($^1$S)+H$_2$ reaction less favorable. Further, the radiative decay of O$^1$S is faster ($A_{5577} > 200A_{6300}$) requiring even higher densities for the reaction with H$_2$ to be competitive. The 5577\AA\ line is therefore not expected to be lowered significantly. In dense regions of disks, the ratio of 6300\AA/5577\AA\  is therefore expected to be lower than $\sim 7$.  
$\dot{N}_{OH}$ photodissociations per second of OH will produce  $\sim 0.5 (1-f_O) \dot{N}_{OH}$ 6300\AA\ photons, where $(1-f_O)$ is the reduction factor due to O($^1$D) + H$_2$ $\rightarrow$ OH which suppresses the radiative transition leading to the 6300\AA\ line. We therefore calculate that the
 [OI]6300\AA\ and [OI]5577\AA\ line luminosities  due to OH photodissociation are $1.1\times10^{-6}$\ls\ and $9.3\times10^{-7}$\ls  in the inner disk respectively, and these lines originate in the same region as the OH cascade, at $r<0.1$AU, where $n\sim 10^{12-13}$cm$^{-3}$ and $T\sim 500-1000$K. OH photodissociation in the mid-disk regions produces line luminosities $5\times10^{-6}$\ls\  (6300\AA) and $7\times10^{-7}$\ls (5577\AA). This emission is produced from gas with $n\sim10^{7-10}$cm$^{-3}$ and $T\sim 200-600$K out to radii $r\sim 40$AU.    Therefore, most of the OI luminosity from TW Hya comes from photodissociation of OH. Ercolano \& Owen (2010) recently propose that the OI forbidden line emission seen in disks may arise from a X-ray driven wind.  Our picture differs from theirs in two ways -- in our model the OI emission arises through a non-thermal process, and the emission arises from the disk atmosphere rather than in a wind.  This perspective is consistent with the OI 6300/5577\AA\  line ratio and the lack of a blueshift for the emission.

The total luminosities from the SML disk (thermal +photodissociation) are therefore $8.4\times10^{-6}$\ls\ for the [OI]6300\AA\ line and $1.6\times10^{-6}$\ls for the [OI]5577\AA\ line. Our 6300/5577\AA\  line ratio is $\sim 6$. The SMH disk which was a better match for the OH emision gives slightly higher [OI] emission, $1.5\times 10^{-5}$\ls\ for the 6300\AA\ line and $1.92\times10^{-6}$\ls\ for the 5577\AA\ line for a ratio of 7.8 (also see Table~\ref{idtable}). 

The intrinsic FWHM linewidth of the observed OI lines are of order 10 km s$^{-1}$ (Pascucci et al. 2011).  This corresponds to a velocity dispersion of about 5 km s$^{-1}$.  In the SML disk, the OI lines arise mostly in the mid-disk regions where the projected Keplerian speeds are low, $< 2$ km s$^{-1}$.  In addition, the gas is relatively cool and turbulent speeds are expected to be low (Hughes et al. 2010).  Hence, the contribution to velocity dispersion from thermal and turbulent velocity fields is expected to also be low, $< 1$ km s$^{-1}$.  However, the photodissociation of OH leads to a recoil of the resultant O($^1$D) or O($^1$S) atoms at speeds of order 2 km s$^{-1}$, and the [OI] photons are emitted prior to collisional relaxation (van Dishoeck \& Dalgarno 1984).  We estimate the linewidth of the OI 6300\AA\  line by adding the various components and the Keplerian, thermal and recoil contributions as appropriate. We obtain a linewidth of $\sim 6$km s$^{-1}$, narrower that the observed value. In the SMH disk, the calculated linewidth is more in agreement with observations, 
$\sim 11$km s$^{-1}$, because of the increased prompt emission close to the star, although the total OI6300\AA\  line luminosity is overproduced by a factor of $\sim 1.5$ (Table~\ref{idtable}). 

We end with a caveat that although the above production mechanism via OH photodissociation  explains the OI6300\AA\ and OI5577\AA\ line strengths, ratio, and lack of an observed blue-shift in TW Hya, it may not be a more general explanation valid for all disks in which optical forbidden line emission has been detected, especially if these lines are blue-shifted and originate in a flow. For disks with different surface density distributions and FUV and X-ray luminosities, there may be a more dominant thermal contribution to [OI] emission. Further, our disk models are static and do not include any emission that might arise in an extended wind (e.g., Ercolano and Owen 2010). Although we expect that our disk density and temperature structure is reasonably accurate within our grid, we do not account for any emission that might arise from heights above our vertical extent of $z=r$. Densities in these regions are typically low, and usually not a significant source of emission. For the optical lines (OI and SII, discussed below) however, the upper states are at $\sim 20,000-40,000K$ and the emission is very sensitive to temperature, and the contribution from the wind in this case might be significant.

\paragraph{SII optical lines}
Pascucci et al. (2011) re-analyze the optical data of Alencar \& Batalha (2002) and report the detection of the [SII]4069\AA\ line ($\sim 3\times10^{-6}$\ls) from TW Hya. They do not detect other lines of SII and place an upper limit of $1.7\times10^{-7}$\ls\ on the 6731\AA\  line. We find that in our SML disk, most of the SII emission arises from the inner disk in hot, dense gas that also results in the thermal OI emission calculated above. Our calculated SII line luminosities are $3.9\times10^{-6}$\ls\ for the 4069\AA\ lines, and $8.6\times10^{-7}, 5.6\times10^{-8}$ and $1.1\times 10^{-7}$\ls\ for   
the undetected lines at 4076\AA, 6716\AA\ and 6731\AA\ respectively. Our [SII]4069\AA\ line is slightly stronger than observed, however, sulfur may be somewhat depleted compared to our assumed abundance($2.8\times10^{-5}$).

\paragraph{NeII and NeIII}
[NeII] emission from TW Hya is strong and has been detected by 
Ratzka et al. (2007), Herczeg et al. (2007), Pascucci \& Sterzik (2009) and by Najita et al. (2010). There is some discrepancy between the measured values, with more flux observed by the {\em Spitzer} study. Najita et al. (2010) conclude that  [NeII] emission arising from a more extended region ($r\gtrsim10$AU) explains their higher flux at all 
three observational epochs, which the narrow slits of the ground-based observations exclude (Pascucci \& Sterzik 2009). A revised calibration (Pascucci et al. 2011)  recovers the {\em Spitzer} flux, and suggests that most of the [NeII] emission comes from within $\sim 10$AU.  Earlier work by Herczeg et al. (2007) measured a line luminosity of $4.4\times10^{-6}$\ls with their ground-based study, similar to the {\em Spitzer} data. 
There is therefore some indication that the [NeII] line flux (or the underlying continuum) might be variable (Pascucci et al. 2011).  Here we use the {\em Spitzer} IRS measurements of Najita et al. (2010) for the [NeII] and [NeIII] emission, $3.9-4.7\times 10^{-6}$L$_{\odot}$ and $2\times 10^{-7}$L$_{\odot}$. 

[NeII] emission from the disk models is found to be quite insensitive to  the surface density distribution. This is because in X-ray heated regions, the [NeII] emission mainly depends only on the intercepted X-ray photon flux and the temperature of the emitting gas (HG09). A simple estimate of the expected [NeII] line luminosity can be made using the analytical expression derived in HG09,  
\begin{equation}
L_{[NeII]}^X \sim  10^{-6} \ \ T^{0.26} e^{-1120/T} \left({f}\over{0.7}\right) \left({\phi_X}\over{10^{39}{\rm s}^{-1}}\right) 
{\rm L}_{\odot}
\end{equation}
where $\phi_X\sim 10^{39}$s$^{-1}$ is the 1keV X-ray photon luminosity. In the above, we also assume that the disk intercepts a fraction $f\sim0.7$ of the stellar X-ray flux (our grid in the numerical work typically extends to $z_{max}=r$). A typical gas temperature in the [NeII] emitting layer is $T\sim2000$K. The expected [NeII] luminosity for the TW Hya disk is therefore, 
$4\times10^{-6}$\ls. 

From our full, detailed numerical calculations, we obtain total [NeII] and [NeIII] line luminosities for the SML disk as  $3\times 10^{-6}$L$_{\odot}$ and $3\times 10^{-7}$L$_{\odot}$ respectively.  Approximately 25\% of the [NeII] comes from gas in the inner disk (see Fig.~\ref{lines}).  The mid-disk region ($3.5{\rm AU}<r<20{\rm AU}$) contributes the remaining 75\% to the [NeII] luminosity in the SML disk. The [NeII] arises from the X-ray heated and partially ionized surface of the disk where gas temperatures range from $1000-4000$K. [NeII] emission from the SML disk is restricted radially to $r\lesssim 10$ AU (Fig.~\ref{lines}), and peaks strongly at the vertically extended 4 AU rim. The X-ray produced [NeII] is somewhat lower than the observed line flux, which could be explained either by the presence of an EUV component and/or by variability of the stellar high-energy photon flux. We note that the accretion rate of TW Hya is observed to be variable on timescales similar to that for the [NeII] line flux (Eisner et al. 2010), $\sim$ a few years. As the X-ray and any [NeII] producing EUV radiation fields may be a result of accretion-induced stellar activity (Robrade \& Schmitt 2006, HG09), this may naturally explain the variability of the [NeII] line. 

Pascucci \& Sterzik (2009) resolve [NeII] in their spectra, and the observed linewidths and blue-shifted emission are consistent with a photoevaporative flow from the star. The line profile is very well reproduced by EUV ($h\nu >$13.6eV) photon-driven photoevaporation models (Alexander 2006). However, we do {\em not} include EUV in our source spectrum for TW Hya and can recover most of the observed flux from {\em X-ray heated gas} alone. The remaining $\sim$25\% of the observed NeII flux may be due to the EUV layer. We can thus estimate the EUV photon luminosity as being $\lesssim 3\times10^{40}$ s$^{-1}$. The [NeII]/[NeIII] line ratio may provide some measure of the origin of the emitting gas (the completely ionized EUV-heated layer or the partially ionized X-ray-heated layer below the EUV-heated layer; $10,000$K in the EUV case and $\sim 1,000-3,000$K for X-rays), but is not definitive because of the uncertain nature of the EUV spectrum (HG09).  However, the most natural explanation is an origin in the X-ray heated gas, where rapid charge exchange of Ne$^{++}$ with H atoms quenches the [NeIII] line and produces the observed [NeII]/[NeIII] ratios. (For  a more detailed discussion, see HG09.) The observed blue-shift may also arise for a X-ray induced photoevaporative flow (Gorti et al. 2009, Owen et al. 2010), and the lack of need for significant EUV to reproduce the [NeII] suggests that any photoevaporative flows may originate in primarily  neutral, X-ray heated gas.

The far-infrared and sub-millimeter emission lines discussed hereon originate only in the mid and outer-disk regions and are the same for the two disk models, SML  and SMH. The line luminosities are listed in Table~\ref{SMLtable}. 

\paragraph{ [OI]  and [CII] fine structure lines}
We obtain a [OI]63$\mu$m line emission of $6\times10^{-6}$\ls, a factor of $\sim 2$ higher than the {\em Herschel} PACS measurement of Thi et al. (2010). Our calculated [OI]145$\mu$m line luminosity is  $ 3\times 10^{-7}$\ls, and consistent with the upper limit of $4\times 10^{-7}$\ls\  obtained by Thi et al. (2010). Emission is extended and produced by $\sim 50-100$K gas at $r\sim30-120 $AU. 

 Our model overproduces [OI]63$\mu$m emission, and we suggest that inclusion of water freezing on grains might alleviate this problem.  Freezing of water ice on grains colder than $100$K removes gas phase elemental oxygen from the outer disk where dust grains are cold. Photodesorption would release some of this ice and at A$_{V}<3$,  all water ice is photodesorbed from grains (Hollenbach et al. 2009). Including these processes reduces the [OI]63$\mu$m emission from the disk compared to a disk with no ice formation, to $4\times10^{-6}$\ls, in better agreement with the PACS observations.  The [OI[145$\mu$m line is reduced from $3\times10^{-6}$\ls \  to $2\times10^{-6}$\ls.   A more detailed treatment of freeze-out of molecules on dust grains and desorption processes will be a subject of future work. Another possibility is that the PAH abundance in the outer disk is even lower than we assume, with a depletion factor of nearly $\sim 1000$. This lowers the temperature in the outer disk sufficiently and decreases the [OI]63$\mu$m line luminosity to $4\times10^{-6}$\ls. If freezing were 
also included, this would bring the OI emission even closer to the observed value.  However, we note that decreasing the PAH abundance everywhere in the disk alters the vertical structure of the mid-disk region ($r<20$AU) sufficiently to deteriorate  fits to observation of other species such as H$_2$, OH and OI forbidden lines. There may, in fact, be a radial gradient in the PAH abundance in the disk, with a higher degree of depletion in the outer disk, if PAHs were to enter the disk frozen out on dust grains as suggested by Geers et al. (2009). Another possibility is that the PAH abundance is low everywhere in the disk, but that other sources of heating such as mechanical viscous dissipation are effective at the surface in the disk($r \lesssim 30$AU) (e.g., Glassgold et al. 2004). 

Thi et al. (2010) model [OI] emission from TW Hya and conclude that the observed flux together with their $^{13}$CO observations indicate a low gas disk mass by factors of $10-100$ compared to that expected from the disk dust mass and assuming a gas/dust ratio of 100.  However, we find that lowering the gas disk mass by 10 and retaining the dust mass only marginally improves model fits to the observed  [OI] emission, and worsens the other line emission fits considerably. Most of these discrepancies can be attributed to the increased grain absorption cross section per H in a disk with a low gas/dust ratio.  H$_2$ emission increases (FUV heating by PAHs becomes significant) by a factor of $\sim 2-3$, OH  prompt emission is reduced due to the increased (relative to gas) absorption of Lyman $\alpha$ photons by dust, and  there is excess emission from CO pure rotational lines. The HCO$^+$4-3 line requires high densities for emission, and this line luminosity is almost a factor of 10 lower for the low mass disk than the observed value.  Increasing the PAH abundance by 10 as in the preferred model of Thi et al. would raise gas temperatures even higher and further increase H$_2$ and CO emission. 
 Although we do not concur with the Thi et al. (2010) results, we note that our calculated model line fluxes are in good agreement with results from other models they consider. Our model  is most similar to their low PAH abundance models (their series 2 and 3) and if they were to base their conclusions on these alternate models, a disk mass close to $\sim 0.06$\ms\  is inferred from their results. For these series, they also seem to find that the OI line luminosity is quite insensitive to disk mass in the range we consider. Their conclusion of a low gas disk mass appears to be largely based on the ratio of the $^{12}$CO/$^{13}$CO line fluxes. In order to accurately interpret this measurement, a more accurate disk chemical model which includes $^{13}$CO as a species is necessary. Our model does not include $^{13}$CO. We also note that Thi et al. assume a constant $^{13}$CO to $^{12}$CO ratio to calculate the $^{13}$CO flux and do not include a proper chemical model of $^{13}$CO as well.  Moreover, the $^{13}$CO/$^{12}$CO line ratio will be significantly affected by processes such as selective photodissociation (Lyons \& Young 2005) and freeze-out processes, which are not included in our disk models or those of Thi et al. (2010). It remains to be seen whether a low gas disk mass is needed to reproduce the observed CO isotope flux ratio.

The calculated [CII]158$\mu$m line luminosity is $3\times 10^{-7}$\ls, and is consistent with the upper limit of $4.5\times10^{-7}$\ls\ set by the {\em Herschel} observations. 

\paragraph{CO rotational lines}
  The  observed line luminosities of the CO 6-5, 3-2 and 2-1 lines are
$5.0\times10^{-8}, 2.5\times10^{-8}$ and $7.9\times10^{-9}$L$_{\odot}$ respectively, whereas we obtain $8\times10^{-8}, 2.4\times10^{-8}$ and $7.9\times10^{-9}$L$_{\odot}$ for these lines. Purely thermal line broadening in the outer disk provides better 
correspondence between data and model results. This is in fair agreement with the results of Qi et al. (2006) who assume only a small amount of turbulence ($\Delta V \sim 0.3$ kms$^{-1}$) in their analysis and with the more recent results of Hughes et al. (2010) who find that the turbulent linewidths are low.  
 A power-law surface density profile also matches the data, but has to be arbitrarily truncated at $\sim 70$AU (GH08). We consider it more realistic to assume that the surface density declines more smoothly with radius, as in a disk evolving viscously and subject to photoevaporation (also see Hughes et al. 2008, Isella at al. 2009).

CO rotational line emission constrains the distribution of gas in the outer disk. We do not model disk masses higher than 0.06\ms, because we expect that appreciably higher disk masses will make the disk gravitationally unstable.  Lower disk masses by a factor of 10 (M$_{disk}=0.006$\ms) do not fit the CO rotational line emission well. We find that although it is possible to match the observed CO 2-1 and 3-2 line emission by altering the surface density profile in the outer disk for the lower mass disk, the CO 6-5 line emission is higher by a factor of $\sim 4$ than the SML value, and hence discrepant with data.  We expect freezing will not affect the emission line luminosities calculated. A simple estimate made by discounting contributions from spatial locations where the dust temperature is $\lesssim 20$K only slightly lowers the calculated line luminosities by $\sim 15\%$. Photodesorption may also keep the CO molecules from freezing on dust grains in the surface regions of the outer disk.  
 We note that we do not need severe radial depletion of CO to explain the observed line ratios (e.g., Qi et al. 2006), although some degree of depletion (less than a factor of 3) in the mid-disk region may help lower the CO 6-5 line luminosity from the model disk. 
 
\paragraph{HCO$^+$ 4-3 line}
van Zadelhoff et al. (2001) measured the line intensity of the HCO$^+$ 4-3 transition from TW Hya using the JCMT, and the line luminosity is $7.7\times10^{-9}$\ls. Our calculation yields  $4\times 10^{-9}$\ls, nearly a factor of 2 lower than observed. We could  obtain a better match by changing the surface density distribution in the outer disk to increase the strength of the  HCO$^+ 4-3$ line, but this then also changes the CO rotational line intensities and makes that agreement poorer. 
 Further, increasing the outer disk surface density also results in a high overall disk mass, bringing the disk closer to being gravitationally unstable.   The HCO$^+$ line mainly originates in relatively warm ($T\sim 20-100$K), and high density gas ($n\gtrsim10^{7-8}$cm$^{-3}$) at $r\lesssim 50$AU.  The abundances of HCO$^+$ calculated by the model in the emitting layers of the disk are a factor of $20-70$ lower than the typical dark cloud abundance of $5\times10^{-9}$ adopted by van Zadelhoff et al. (2001) in their analysis, and thus consistent with their derived ``depletion" factors of $10-100$.

We conclude this section with the caveat that our models may not be unique solutions to the gas surface density distribution in the disk. There are theoretical uncertainties
 that are inherent in chemical rates and the microphysics, as well as several important processes (such as freezing of ices) have been either simplified or ignored.  The complicated and computationally intensive modeling procedure  also makes a full exploration of parameter space  impossible at this time.    Nevertheless, we find it promising that theoretical models with reasonable parameters can reproduce observed emission ranging from the optical to the sub-millimeter wavelengths with an accuracy to within a factor $\lesssim$ 2.

\section{The Origin of the Inner Gas Hole}
Modeling of gas emission lines and comparisons with observational data indicate that the surface density of gas is also depleted in the inner $r<4$AU dust depleted region. The models are  further consistent with a massive, optically thick outer disk beyond the 4AU rim.  At least three different explanations have been proposed to explain the dust holes in transition disks, of which TW Hya is a leading example (e.g., Najita et al. 2007). The distribution of gas in the disk that we derive may help distinguish between these various scenarios. We examine the implications of each of the explanations for transition disks and discuss our model results in each context.
\paragraph{Grain growth with no gas depletion in the dust hole:}
The first step towards planet formation in disks via the core accretion scenario is the coagulation of   primordial sub-micron sized dust grains into larger dust particles and
planetesimals. Grain growth is expected to be more efficient in the inner disk as the densities there are higher (e.g., Lissauer et al. 2007). The population of small dust decreases due to grain growth, and as larger solids are not efficient at absorbing starlight because of their lower collective surface area,  this process is manifested as  a decrease in the dust continuum emission at shorter wavelengths where the warm small dust in the inner disk would emit. Grain growth does not remove disk gas, and no depletion in gas surface density within the inner regions is expected (unless gas giant planets have already formed, which is the next scenario).   In this case, there would be continued accretion onto the star (as is observed for TW Hya). 

 Our modeling of the gas disk around TW Hya when compared with observational data does not support the grain growth explanation alone as a mechanism to cause  the dust hole. Based on the constraints set by gas line and continuum emission, we argue that gas must also be depleted in the  inner disk region ($r\lesssim4$AU), with  a surface density decrease $\sim 1-2$ orders of magnitude compared to what one would obtain by extrapolating the surface density from  the outer optically thick, massive disk. Although grain growth may explain the decrease in dust opacity inside 4 AU, a mechanism that depletes gas surface density by $\sim 10-100$ is also necessary.

\paragraph{Giant Planet Formation} 
A widely favored explanation for the observed dust depletion in transition disks is the creation of a gap by a giant planet, followed by the formation of a hole as the inner disk accretes onto the star.   A gas giant planet exerts dynamical torques on the disk and clears material in its immediate vicinity to create a gap (e.g. Artymowicz \& Lubow 1996, Bate et al. 2003, Rice et al. 2003). The disk interior to the planet is expected to rapidly accrete onto the central star. The outer disk mainly accretes onto the planet, with a small fraction of the mass flow (depending on the ratio of planet mass to disk mass) continuing to accrete onto the star (Lubow and D'Angelo 2006).   Pressure maxima in the disk just outside the planet may act to filter dust particles (Rice et al. 2006) and further deplete the inner disk of dust. Our modeled surface density distribution supports this explanation for the TW Hya disk, with the potential planet located between $2-4$AU  where the surface density may be quite low (with gas in accretion streams)  and consistent with the presence of a gap.  The accretion streams then feed a depleted inner accretion disk at $\lesssim 2$AU. 

We can use the surface density distribution determined by the gas emission models to set constraints on the mass of the planet orbiting TW Hya.  
As the planet must open a gap in the disk, or at least establish a large surface density contrast at its location, we can use the gap-opening criterion to obtain a minimum mass $M_p$ of the planet (e.g., Lin \& Papaloizou 1986) 
\begin{equation}
\left({{M_p}\over{M_*}}\right)^2 \gtrsim 3 \pi \alpha \left({{H}\over{r}}\right)^5
\end{equation}
where $H$ is the scale height of the outer disk at the radius $r\sim 4$ AU. From our disk models, $H/r \sim 0.04$ at 4AU, and therefore the planet has a minimum mass of $\sim 0.03-0.3M_J$ for $\alpha\sim10^{-3}-10^{-1}$ in the outer disk. Simultaneously, one can appeal to recent hydrodynamical models of accretion onto and past a massive planet (Lubow \& D'Angelo 2006), which find that a planet 
with at least this minimum mass must accrete matter at a rate $\sim 10$ times higher than the rate onto the star. Given the current age of the system  ($\sim$10 Myrs), the planet is likely to have undergone accretion for at least 1 Myr. Even with the lower estimate of the stellar accretion rate $\sim 4 \times 10^{-10}$\ms\ yr$^{-1}$, this implies a minimum planet mass $M_p \sim 4M_J$. An upper limit to the planet mass can also be estimated from the stellar accretion rate. The fact that gas in the outer disk can still flow through the gap induced by the planet suggests the planet is not massive enough to form a tidal barrier and repel disk matter. Depending on $\alpha$, this limits the planet mass to $\sim 7 M_J (\alpha=4\times 10^{-3})$ or $ \sim 2 M_J (\alpha=4\times10^{-4})$ (see  D'Angelo et al. 2010). 
We conclude that a plausible planet mass is likely to be $\sim 4-7 M_J$.  

 The mass of the planet inferred above is high enough that dust particles with sizes larger than $\sim 1\mu$m may be trapped in the pressure maximum in the disk at the outer edge of the gap (Rice et al. 2006). Dust particles with sizes of $\sim 2\mu$m have been modeled to exist in the inner disk (C02), consistent with this scenario. Since the inferred dust distribution in the outer disk has a maximum size of $a_{max}=1$cm, and only sub-micron sized grains cross the gap, dust filtering  would result in a decrease in dust mass inside the gap.   Therefore an increase in the gas/dust mass ratio across the gap by a factor of $\sim 100$ (since the dust mass is proportional to $\sqrt(a_{max})$) is expected, which is consistent with our SMH model disk.

A potential issue with the planet hypothesis is the migration rate.  For this case, we need to consider Type II migration as the planet must be massive enough to create a gap. Viscous diffusion of the disk causes the planet to drift towards the star efficiently, but only as long as the local disk mass ($\pi r^2 \Sigma(r)$, $r\sim4$AU) is greater than the planet mass.  For the mid and outer disk surface density distribution, the local disk mass is $\sim 1M_J$. Thus the expected planet of $\sim 4M_J$ is stable against migration.

\paragraph{Photoevaporation\\}
The strong FUV and X-ray flux from TW Hya should drive massive photoevaporative flows from the disk (Ercolano et al. 2008, Gorti \& Hollenbach 2009, Gorti et al. 2009, Owen et al. 2010).  Photoevaporating flows have been observed by resolved ground-based [NeII]12.8$\mu$m emission from TW Hya (Pascucci \& Sterzik 2009). The observed blue-shifted line profiles are well-reproduced by theory (Alexander 2006, Owen \& Ercolano 2010).

 However, whether the inner hole in TW Hya  is carved by photoevaporation remains a question. The location of the radius of the hole $\sim 4$ AU is consistent with theory as it is larger than the so-called critical radius, $\sim 1$AU where the hole first appears in photoevaporative theory before the disk is eroded radially outward by direct rim illumination.  However, photoevaporation theory traditionally predicts that a ``clean" hole, completely devoid of dust and gas, must rapidly form on viscous timescales (e.g., Alexander et al. 2006).  
 The viscous clearing timescale in the inner disk is $t_{vis}\sim r^2/\nu$, or $t_{vis} \sim \sqrt{(GM_* r)}/(\alpha c_s^2)$. For  the TW Hya disk,  $t_{vis} \sim 10^4$ years at 4AU for $\alpha \sim 0.01$. Therefore, the surface density drop of $\sim 100$ at 4 AU implies an e-folding factor of 5, or $\sim 5\times10^4$ years after gap opening by photoevaporation. 
 For a higher surface density in the inner disk and a drop of $\sim 10$ at 4 AU, gap opening must be even more recent. 
Although there must be ongoing photoevaporation from the disk as is observed, our results appear to refute current photoevaporation theories for the creation of the inner hole in TW Hya, {\em unless} the gap creation epoch is extremely recent, which seems unlikely
 given the $\sim 10^7$ year age of TW Hya. 

We note that one possibility where one could avoid a special observing epoch and where photoevaporation could work as a gap creation mechanism would be if the thermal photoevaporative flow that emanates from the rim and disk surface is partially re-captured by stellar gravity to form the inner disk and eventually accrete onto the star. In our models, the photoevaporative mass loss rate is $\sim 4\times10^{-9}$\ms yr$^{-1}$.  For a stellar accretion rate of   $\sim 4\times10^{-10}$\ms\ yr$^{-1}$, 10\% of the flow needs to be recaptured by the star.  We use simple timescale arguments to show that this is a feasible process. For recapture, the flow timescale ($t_f$) to the sonic radius $r_s$ where gas escapes needs to be comparable to the viscous timescale $t_{vis}$ of the photoevaporating gas.  As $t_f \sim r_s/c_s$, and using $r_s=r_g/2$ and the expression for $t_{vis}$ in the preceding  paragraph, we obtain $t_f/t_{vis} \sim \alpha (r_g/r)^{1/2}$.  For FUV/X-ray driven photoevaporation where the flow is launched subsonically (GH09, GDH09) from the rim at $r=4$AU, and a typical gas temperature of $\sim 2000$K, $r_g =31$AU. For 10\%  of the flow is to be recaptured due to viscosity, we need $t_f/t_{vis}\sim 0.1$ or a value of the viscous parameter $\alpha=0.1$.   If the accretion rate is $\sim 10^{-9}$\ms yr$^{-1}$, the mass captured needs to be $\sim 25\%$ for the calculated photoevaporation rate.  Therefore, for this re-capture process to be possible,  $\alpha$ needs to be relatively high in the inner disk, $\sim0.1-0.25$.  However,  this process is a complicated phenomenon that requires  full 2D  hydrodynamical calculations to resolve, including turbulent viscosity in the flow that can redistribute the angular momentum.   This problem will be the subject of future study.

We also note that even if the inner opacity hole is not created by photoevaporation, the process must still act to disperse the disk. The observed [NeII] line profiles indicate this is an ongoing process in the TW Hya disk.  We can estimate the lifetime of the remaining disk around TW Hya using FUV/X-ray  photoevaporation rates from theory (GH09) and the results from the SML disk.  From our calculated disk photoevaporation rate $4\times10^{-9}$\ms\ yr$^{-1}$, we derive a disk lifetime of $M_{disk}/\dot{M}_{pe}$ as $\sim 10^7$ years. Note that with the reduced disk mass of Thi et al., the expected disk lifetime is $\lesssim 10^6$years, which places us in an unlikely viewing epoch
and therefore seems implausible.  We calculated a time evolution model of the disk, with the SML disk as an initial configuration and assuming a viscosity parameter $\alpha=0.005$ in the outer disk using the photoevaporation models described in Gorti et al. (2009). When the effects of viscous evolution are included, the remaining lifetime of the disk is calculated to be $5\times10^{6}$ years. We note that a higher assumed value of $\alpha$ would lead to even shorter disk lifetimes. 

Most of the photoevaporation occurs at the 4AU rim, although the mass loss rate derived is lower compared to that from holed disks with directly illuminated rims (e.g., Gorti et al. 2009, Owen et al. 2010) because of the partial shielding of the rim in this case by gas in the inner disk. 
The photoevaporation rate is higher than recent measurements of the stellar accretion onto TW Hya, which range from $4\times10^{-10}-3\times10^{-9}$\ms\ yr$^{-1}$
(Muzerolle et al. 2000, Alencar \& Batalha 2002, Eisner et al. 2010). This is consistent with the requirement for photoevaporation to create a gap in the disk, when the accretion rate at $\sim 1-4$AU must drop below the photoevaporative mass loss rate.

 If photoevaporation has caused the inner hole and not a giant planet, then the amount of gas we infer to be present in the inner disk and the observed accretion onto the star imply  that either the gap has been very recently created ($\lesssim 10^5$ years ago), or, that the gap was created earlier and that the observed accretion is caused by viscosity in the flow causing significant re-capture of the photoevaporating gas and accretion onto the star.  If a giant planet has caused the inner hole, then the accretion rate in the outer disk is expected to be at least a factor of 10 higher than $\dot{M}_*$, and in this case the derived photoevaporative mass loss rate is lower than the outer accretion rate, which is consistent with the giant planet hypothesis and, in this case, not consistent with the pure photoevaporation scenario for hole formation.

\section{Constraints on the Disk Mass from  Line Emission Modeling}
In this paper, we model the gas disk of TW Hya and conclude that the best model which reproduces the observed gas emission line luminosities is one where there is an optically thick disk ($0.06$\ms) extending from $\sim$3.5AU outwards, with a depleted inner region that contains  $\sim 10^{-4}- 10^{-5}$\ms\  of gas  extending to the star. The modeling suggests a lower gas mass in the inner disk by $\sim 10-100$ than obtained by extrapolating the surface density of the outer disk radially inward.   Most of the disk mass ($0.06$\ms) is contained within 100AU, and the surface density in the outer disk declines sharply beyond this radius.  {\ The gas/dust mass ratio is consistent with the canonical ratio of 100 in the outer optically thick disk, and is enhanced by factors of $\sim 5-50$ in the inner disk where both gas and dust are severely depleted compared with the extrapolated surface density of the outer disk.   Although the outer disk gas/dust ratio is the same as the interstellar ratio, the cross sectional area of dust per H nucleus is $\sim 100$ times less than interstellar dust, because the disk dust size distribution extends to much larger sizes, $a_{max}\sim 1$mm, than interstellar dust where $a_{max}\sim0.2\mu$m.   Therefore there are far fewer small dust particles (which
provide most of the cross sectional area) per H nucleus, since much of the dust mass is tied up in the largest size ($\sim 1$ mm) particles.

\paragraph{Inner Disk} The observed CO rovibrational emission and limits on H$_2$O line emission between 10-19 $\mu$m
constrain the inner disk mass, and are best explained by a  gas mass  $\sim 10^{-5}$ \ms\  within $r<3.5$AU. Lower disk masses ($10^{-6}$\ms) are very discrepant with  observed high J CO rovibrational emission and can be ruled out. A full, undepleted disk with mass $\sim 10^{-3}$\ms\ is rejected based on the expected continuum emission due to gas opacity and the lack of observed mid-IR water emission. A disk mass of $10^{-4}$\ms\ is, however, also in reasonable agreement with data. 
 There is some uncertainty in the inner disk dust mass  as inferred from observations ($\sim 2\times 10^{-8}$\ms,C02;  $\sim 10^{-10}$\ms, Eisner et al. 2006), but the conclusion that the hole is depleted of gas and dust compared to the outer disk appears quite robust.

The amount of mass present in the inner disk is fairly consistent with the observed accretion rate. Assuming  steady state accretion rate onto the star $\dot{M}_{*}=3 \pi \nu \Sigma$, where the viscosity 
$\nu= \alpha c_s^2/\Omega$, $\alpha$ being the viscosity parameter, $c_s$ the sound speed of gas and $\Omega$ the angular velocity (e.g., Pringle 1981), we can estimate $\alpha$ from the observed $\dot{M}_*$ and from $\Sigma$ and $c_s$ derived from our models. For an accretion rate $\dot{M}_{*}=4\times10^{-10}$\ms\ yr$^{-1}$, and at $r=1$AU, we then obtain $\alpha \sim 0.04$ for the SML disk and $\sim 0.004$ for the SMH disk. (If  $\dot{M}_{*}$ were $\sim 10^{-9}$\ms\ yr$^{-1}$, then $\alpha=0.1$ for the SML disk and $\alpha=0.01$ for the SMH disk.) 
 At the inner edge, close to the magnetospheric truncation radius of $0.06$AU, $\alpha$ is higher, $\sim 0.08-0.008$. Our derived values for $\alpha$ for the preferred SML disk are  somewhat higher than the $0.01-0.005$ value typically used for protostellar disks, but consistent with the higher gas temperatures and ionization levels expected in the inner disk ( King et al. 2007, Turner et al. 2010).  For the SMH  inner disk mass of $10^{-4}$\ms, the inferred value of $\alpha$ is more in line with values typically used.

\paragraph{Outer Disk} HCO$^+$ and CO rotational lines place upper limits on the amount of gas that is present in the outer disk, and are consistent with a disk gas mass $\sim 0.06$\ms. All the other observed line emission originates mainly in the disk at radii $r\lesssim30$AU, which contains about 30\%\ of the total disk mass.  We find that decreasing the gas mass by a factor of 10 lowers the HCO$^+$ luminosity  by a factor of 5 due to its origin in dense gas.  Since our model already underestimates this luminosity by a factor of 2, this would produce a major discrepancy in the model luminosity versus the observed HCO$^+$ luminosity. In the adopted surface distribution with a tapering outer edge, CO pure rotational lines are also quite sensitive to the mass in the outer disk. The lower mass disk results in a high CO 6-5/2-1  line ratio of 16, compared with the observed ratio of 5.8 (see Table~\ref{obsdata}). Increasing the gas mass by a factor of 10 brings the disk mass to nearly the stellar mass, and such a disk would be highly gravitationally stable (Pringle 1981). We therefore consider a more massive disk than 0.06\ms\  very unlikely.  In the outer disk, gas line emission may therefore constrain disk gas mass to within a factor of $\lesssim 10$, even though most of the gas line emission originates in the surface layers which do not trace the bulk of the disk mass.

We can also set  limits on the disk mass from arguments based on the presence of a planet and from expected mass loss rates due to photoevaporation. If a planet were present and accreting at $\sim10$ times the stellar accretion rate, this would imply a relatively short disk survival time for a 0.006\ms\  disk. Even for a low accretion rate onto the star of  $\dot M_* = 4\times10^{-10}$\ms\ yr$^{-1}$, the disk survival time would be $\sim 6\times10^{-3}$\ms/$(10 \dot{M}_*) \sim$ 1Myr, which is short compared to the age of the system. This situation is  made even more dire if the disk mass were lower or if the stellar accretion rate is at the higher range of measurements. Further, if the disk is losing mass at the estimated photoevaporation rate of $4\times10^{-9}$\ms\ yr$^{-1}$,  a $0.006$\ms\ disk would dissipate in $\lesssim $ 1Myr, which seems unlikely given the age of TW Hya.  In view of all the above, our preferred gas disk mass is 0.06\ms. 
 
\section{Conclusions}
We infer the gas surface density distribution of the transition disk around TW Hya by modeling  gas line emission from the disk, including H$_2$, HCO$^+$ and CO rotational lines, CO rovibrational lines, OH line emission, [NeII], [NeIII], [OI] fine structure emission and optical OI and SII forbidden line emission. Our  standard best fit model (SML) disk  reproduces observed line emission well to within a factor of $\sim 2$ for most lines, except OH emission which is discrepant by a factor of $\sim 4$.  We consider prompt emission following photodissociation of water as mechanism to produce OH.   If we can increase our gas phase water production at the disk surface by a factor of a few, we can bring all emission lines in the model to within a factor of 2 of the observations.  Vertical advection of icy grains in the mid disk region to surface layers where the ice mantles are photodesorbed may be the most likely source of the extra water vapor.  
We further speculate that photodissociation of OH itself may result in re-formation to a rotationally excited state in very dense regions, which might lead to the observed OH emission. The alternate model of a higher inner disk mass (SMH) provides good matches to the observed emission as well. The SMH disk is better at reproducing the OH emission (prompt) and the [OI]6300\AA\ linewidth, although the CO rovibrational line emission is a poorer fit and some emission lines are over-produced. Our main results on gas line emission are as follows. 
\begin{enumerate}
\item The model disk that best matches the observational data indicates a significant decrease in  gas surface density in the inner opacity hole that causes a dip in the mid-infrared continuum. Our best  models have a gas mass of $10^{-4}-10^{-5}$\ms\  at $0.06{\rm AU}<r<3.5{\rm AU}$, and $0.06$\ms between $3.5{\rm AU}<r<200{\rm AU}$.  We exclude a full undepleted disk with $10^{-3}$\ms\ and  inner disk masses of $10^{-6}$\ms \ and below. The gas/dust mass ratio in the inner disk is inferred to be greater by a factor of $\sim5-50$  than the canonical value of 100 which applies in the outer disk. 
 
\item CO rovibrational emission and upper limits on H$_2$O  rotational lines provide good constraints on the mass in the inner disk. The observed CO rovibrational emission is modeled to arise from hot ($T\sim 500-1000$K) gas inside $\sim0.5$AU, from disk surface regions where hydrogen is predominantly atomic. 

\item  The mass of the outer disk is constrained to lie within a factor of a few of 0.06 \ms.  Significantly higher masses lead to gravitational instability.  Significantly lower masses lead to very poor fits to the CO rotational spectrum and to the HCO$^+$ J= 4-3 line luminosity.
In addition, given the relatively rapid accretion and photoevaporation of the outer disk, lower disk masses give unreasonably short outer disk lifetimes ($<1$Myr).

\item We propose that  OH mid-infrared emission originates from prompt emission following photodissociation of water in the disk very close to the star ($r\lesssim0.1$AU) and from the mid-disk regions ($r\sim 4-40$AU).  Our model underestimates the production of OH by a factor of 4, and we speculate that vertical mixing may bring more water to the surface to be photodissociated.   Grain formation of water ice followed by photodesorption could also increase the water vapor at the surface and thereby also increase the OH emission.  However, enhanced water production might also lead to enhanced [OI]6300\AA\ forbidden line emission by a factor of $\sim 3$ above the observed value. We suggest another promising mechanism for producing OH emission in dense regions where the photodissociation of ground state OH leads to an excited O$^1$D atom that reacts with H$_2$ to form rotationally excited OH that cascades giving rise to the observed MIR lines. 

\item We argue that the forbidden lines of [OI] at 6300\AA\ and 5577\AA\  arise due to the  photodissociation of OH mainly from the mid-disk regions ($r\sim 4-20$AU). Such an origin can successfully explain the  OI6300\AA/5577\AA\ ratio of $\sim 7$ observed. 
About $\sim20$\%  of the [OI]6300\AA \ emission is thermal in nature and has equal contributions from the inner disk ($r\lesssim4$AU) and the outer disk ($r\sim 4-10$AU).  

\item NeII emission may not require the presence of a significant EUV flux from TW Hya and can be explained by X-ray heating of gas. If EUV is present, we estimate a photon luminosity of $\lesssim 3\times10^{40}$s$^{-1}$ in order to explain the slight deficit in X-ray produced NeII line fluxes. 

\item H$_2$ pure rotational line emission is extended and is mainly from radii $r\sim4-30$AU.  [OI]63$\mu$m emission, CO pure rotational lines and HCO$^+4-3$ line trace the outer disk at $r\sim 30-120$AU.

\end{enumerate}
We conclude that grain growth alone cannot be responsible for the dust hole in the TW Hya disk because  gas line emission modeling indicates a factor of $\sim 10-100$ depletion in the gas surface density as well.  The transition disk morphology of TW Hya may be explained by the presence of a $\sim4-7$M$_J$ planet at $r\sim2-3$AU that causes the change in surface density distribution near this region. The disk is presently photoevaporating with an estimated mass loss rate of
 $4\times10^{-9}$\ms yr$^{-1}$. If there were no planet, then photoevaporation alone may be in the process of developing a hole. The calculated photoevaporation rate is greater than the stellar  accretion rate  suggesting that  photoevaporation could have created the gap. However ongoing stellar accretion at current rates poses problems for present photoevaporation theories. Regardless of the hole-creation mechanism, the remaining disk lifetime due to photoevaporation and viscous evolution is estimated to be $\sim$ 5 million years.  
 
\acknowledgments
We thank Suzan Edwards, Gennaro D'Angelo, Kees Dullemond, David Neufeld, Ewine van Dishoeck, Roman Krems, Colette Salyk and Sean Brittain for helpful discussions. We are especially grateful to John Black for alerting us about the reactive nature of excited O atoms. U.G and D.H acknowledge support by a grant under the NASA Astrophysical Data Analysis Program which enabled this work. 
 
\thebibliography

\bibitem[Artymowicz \& Lubow(1996)]{1996ApJ...467L..77A} Artymowicz, P., \& Lubow, S.~H.\ 1996, \apjl, 467, L77 

\bibitem[Acke et 
al.(2005)]{2005A&A...436..209A} Acke, B., van den Ancker, M.~E., \& Dullemond, C.~P.\ 2005, \aap, 436, 209 

\bibitem[Alencar 
\& Batalha(2002)]{2002ApJ...571..378A} Alencar, S.~H.~P., \& Batalha, C.\ 2002, \apj, 571, 378 

\bibitem[Alexander et al.(2005)]{2005MNRAS.358..283A} Alexander, R.~D., 
Clarke, C.~J., \& Pringle, J.~E.\ 2005, \mnras, 358, 283 

\bibitem[Alexander et al.(2006)]{2006MNRAS.369..229A} Alexander, R.~D., 
Clarke, C.~J., \& Pringle, J.~E.\ 2006, \mnras, 369, 229 

\bibitem[Alexander et al.(2006)]{2006MNRAS.369..216A} Alexander, R.~D., 
Clarke, C.~J., \& Pringle, J.~E.\ 2006, \mnras, 369, 216

\bibitem[Alexander(2008)]{2008MNRAS.391L..64A} Alexander, R.~D.\ 2008, 
\mnras, 391, L64 

\bibitem[Bakes 
\& Tielens(1994)]{1994ApJ...427..822B} Bakes, E.~L.~O., \& Tielens, A.~G.~G.~M.\ 1994, \apj, 427, 822

\bibitem[Bary et al.(2003)]{2003ApJ...586.1136B} Bary, J.~S., Weintraub, 
D.~A., \& Kastner, J.~H.\ 2003, \apj, 586, 1136 

\bibitem[Bate et al.(2003)]{2003MNRAS.341..213B} Bate, M.~R., Lubow, S.~H., 
Ogilvie, G.~I., \& Miller, K.~A.\ 2003, \mnras, 341, 213 

\bibitem[Bethell 
\& Bergin(2009)]{2009Sci...326.1675B} Bethell, T., \& Bergin, E.\ 2009, Science, 326, 1675 

\bibitem[Bergin et al.(2003)]{2003ApJ...591L.159B} Bergin, E., Calvet, N., 
D'Alessio, P., \& Herczeg, G.~J.\ 2003, \apjl, 591, L159 

\bibitem[Bitner et al.(2008)]{2008ApJ...688.1326B} Bitner, M.~A., et al.\ 
2008, \apj, 688, 1326

\bibitem[Calvet et al.(2002)]{2002ApJ...568.1008C} Calvet, N., D'Alessio, 
P., Hartmann, L., Wilner, D., Walsh, A., 
\& Sitko, M.\ 2002, \apj, 568, 1008 

\bibitem[Carr 
\& Najita(2008)]{2008Sci...319.1504C} Carr, J.~S., \& Najita, J.~R.\ 2008, Science, 319, 1504 

\bibitem[Cazaux 
\& Tielens(2004)]{2004ApJ...604..222C} Cazaux, S., \& Tielens, A.~G.~G.~M.\ 2004, \apj, 604, 222

\bibitem[Chiang 
\& Goldreich(1997)]{1997ApJ...490..368C} Chiang, E.~I., \& Goldreich, P.\ 1997, \apj, 490, 368 

\bibitem[Ciesla 
\& Cuzzi(2006)]{2006Icar..181..178C} Ciesla, F.~J., \& Cuzzi, J.~N.\ 2006, \icarus, 181, 178 

\bibitem[Currie et al.(2009)]{2009ApJ...698....1C} Currie, T., Lada, C.~J., 
Plavchan, P., Robitaille, T.~P., Irwin, J., 
\& Kenyon, S.~J.\ 2009, \apj, 698, 1

\bibitem[D'Angelo et al.(2010)]{2010arXiv1006.5486D} D'Angelo, G., Durisen, 
R.~H., \& Lissauer, J.~J.\ 2010, arXiv:1006.5486  (To appear in {\em Exoplanets}, ed. S.~Seager, University of Arizona Press, Tucson, AZ)

\bibitem[Dullemond et al.(2001)]{2001ApJ...560..957D} Dullemond, C.~P., 
Dominik, C., \& Natta, A.\ 2001, \apj, 560, 957 

\bibitem[Dullemond 
\& Dominik(2005)]{2005A&A...434..971D} Dullemond, C.~P., \& Dominik, C.\ 2005, \aap, 434, 971 

\bibitem[Eisner et al.(2006)]{2006ApJ...637L.133E} Eisner, J.~A., Chiang, 
E.~I., \& Hillenbrand, L.~A.\ 2006, \apjl, 637, L133 

\bibitem[Eisner et al.(2010)]{2010ApJ...722L..28E} Eisner, J.~A., Doppmann, 
G.~W., Najita, J.~R., McCarthy, D., Kulesa, C., Swift, B.~J., 
\& Teske, J.\ 2010, \apjl, 722, L28 

\bibitem[Ercolano et al.(2008)]{2008ApJ...688..398E} Ercolano, B., Drake, 
J.~J., Raymond, J.~C., \& Clarke, C.~C.\ 2008, \apj, 688, 398 

\bibitem[Ercolano 
\& Clarke(2010)]{2010MNRAS.402.2735E} Ercolano, B., \& Clarke, C.~J.\ 2010, \mnras, 402, 2735 

\bibitem[Ercolano \& Owen(2010)]{2010MNRAS.406.1553E} Ercolano, B., \& Owen, J.~E.\ 2010, \mnras, 406, 1553

\bibitem[Festou 
\& Feldman(1981)]{1981A&A...103..154F} Festou, M., \& Feldman, P.~D.\ 1981, \aap, 103, 154 

\bibitem[Figueira et al.(2010)]{2010EAS....42..125F} Figueira, P., Pepe, 
F., Santos, N.~C., Melo, C.~H.~F., Bonfils, X., Hu{\'e}lamo, N., Udry, S., 
\& Queloz, D.\ 2010, EAS Publications Series, 42, 125 

\bibitem[Geers et 
al.(2006)]{2006A&A...459..545G} Geers, V.~C., et al.\ 2006, \aap, 459, 545 

\bibitem[Geers et 
al.(2009)]{2009A&A...495..837G} Geers, V.~C., van Dishoeck, E.~F., Pontoppidan, K.~M., Lahuis, F., Crapsi, A., Dullemond, C.~P., \& Blake, G.~A.\ 2009, \aap, 495, 837 

\bibitem[Glassgold et al.(1997)]{1997ApJ...480..344G} Glassgold, A.~E., 
Najita, J., \& Igea, J.\ 1997, \apj, 480, 344 

\bibitem[Glassgold et al.(2004)]{2004ApJ...615..972G} Glassgold, A.~E., 
Najita, J., \& Igea, J.\ 2004, \apj, 615, 972 

\bibitem[Glassgold et al.(2009)]{2009ApJ...701..142G} Glassgold, A.~E., 
Meijerink, R., \& Najita, J.~R.\ 2009, \apj, 701, 142 

\bibitem[Gorti et al.(2009)]{2009ApJ...705.1237G} Gorti, U., Dullemond, 
C.~P., \& Hollenbach, D.\ 2009, \apj, 705, 1237 

\bibitem[Gorti  \& Hollenbach(2009)]{2009ApJ...690.1539G} Gorti, U., \& Hollenbach, D.\ 2009, \apj, 690, 1539 

\bibitem[Gorti  \& Hollenbach(2008)]{2008ApJ...683..287G} Gorti, U., \& Hollenbach, D.\ 2008, \apj, 683, 287 

\bibitem[Gorti  \& Hollenbach(2004)]{2004ApJ...613..424G} Gorti, U., \& Hollenbach, D.\ 2004, \apj, 613, 424 

\bibitem[Habart et 
al.(2003)]{2003A&A...397..623H} Habart, E., Boulanger, F., Verstraete, L., Pineau des For{\^e}ts, G., Falgarone, E., \& Abergel, A.\ 2003, \aap, 397, 623 

\bibitem[Harich et al.(2000)]{2000JChPh.11310073H} Harich, S.~A., Hwang, 
D.~W.~H., Yang, X., Lin, J.~J., Yang, X., 
\& Dixon, R.~N.\ 2000, \jcp, 113, 10073

\bibitem[Hartigan et al.(1995)]{1995ApJ...452..736H} Hartigan, P., Edwards, 
S., \& Ghandour, L.\ 1995, \apj, 452, 736

\bibitem[Hollenbach 
\& McKee(1979)]{1979ApJS...41..555H} Hollenbach, D., \& McKee, C.~F.\ 1979, \apjs, 41, 555

\bibitem[Hollenbach et al.(2009)]{2009ApJ...690.1497H} Hollenbach, D., 
Kaufman, M.~J., Bergin, E.~A., \& Melnick, G.~J.\ 2009, \apj, 690, 1497

\bibitem[Hollenbach  \& Gorti(2009)]{2009ApJ...703.1203H} Hollenbach, D., \& Gorti, U.\ 2009, \apj, 703, 1203

\bibitem[Herczeg et al.(2002)]{2002ApJ...572..310H} Herczeg, G.~J., Linsky, 
J.~L., Valenti, J.~A., Johns-Krull, C.~M., 
\& Wood, B.~E.\ 2002, \apj, 572, 310 

\bibitem[Herczeg et al.(2004)]{2004ApJ...607..369H} Herczeg, G.~J., Wood, 
B.~E., Linsky, J.~L., Valenti, J.~A., 
\& Johns-Krull, C.~M.\ 2004, \apj, 607, 369 

\bibitem[Herczeg et al.(2007)]{2007ApJ...670..509H} Herczeg, G.~J., Najita, 
J.~R., Hillenbrand, L.~A., \& Pascucci, I.\ 2007, \apj, 670, 509 

\bibitem[Hu{\'e}lamo et 
al.(2008)]{2008A&A...489L...9H} Hu{\'e}lamo, N., et al.\ 2008, \aap, 489, L9 

\bibitem[Hughes et al.(2007)]{2007ApJ...664..536H} Hughes, A.~M., Wilner, 
D.~J., Calvet, N., D'Alessio, P., Claussen, M.~J., 
\& Hogerheijde, M.~R.\ 2007, \apj, 664, 536 

\bibitem[Hughes et al.(2008)]{2008ApJ...678.1119H} Hughes, A.~M., Wilner, 
D.~J., Qi, C., \& Hogerheijde, M.~R.\ 2008, \apj, 678, 1119 

\bibitem[Hughes et al.(2009)]{2009ApJ...704.1204H} Hughes, A.~M., Wilner, 
D.~J., Cho, J., Marrone, D.~P., Lazarian, A., Andrews, S.~M., 
\& Rao, R.\ 2009, \apj, 704, 1204 
\bibitem[Hughes et al.(2011)]{2011ApJ...727...85H} Hughes, A.~M., Wilner, 
D.~J., Andrews, S.~M., Qi, C., \& Hogerheijde, M.~R.\ 2011, \apj, 727, 85 

\bibitem[Ingleby et al.(2009)]{2009ApJ...703L.137I} Ingleby, L., et al.\ 
2009, \apjl, 703, L137 

\bibitem[Isella et al.(2009)]{2009ApJ...701..260I} Isella, A., Carpenter, 
J.~M., \& Sargent, A.~I.\ 2009, \apj, 701, 260

\bibitem[King et al.(2007)]{2007MNRAS.376.1740K} King, A.~R., Pringle, 
J.~E., \& Livio, M.\ 2007, \mnras, 376, 1740 

\bibitem[Kirby-Docken 
\& Liu(1978)]{1978ApJS...36..359K} Kirby-Docken, K., \& Liu, B.\ 1978, \apjs, 36, 359 

\bibitem[Lin 
\& Papaloizou(1986)]{1986ApJ...309..846L} Lin, D.~N.~C., \& Papaloizou, J.\ 1986, \apj, 309, 846 

\bibitem[Lin 
\& Guo(2008)]{2008JChPh.129l4311L} Lin, S.~Y., \& Guo, H.\ 2008, \jcp, 129, 124311 

\bibitem[Lissauer 
\& Stevenson(2007)]{2007prpl.conf..591L} Lissauer, J.~J., \& Stevenson, D.~J.\ 2007, Protostars and Planets V, 591

\bibitem[Lubow 
\& D'Angelo(2006)]{2006ApJ...641..526L} Lubow, S.~H., \& D'Angelo, G.\ 2006, \apj, 641, 526

\bibitem[Lyons 
\& Young(2005)]{2005Natur.435..317L} Lyons, J.~R., \& Young, E.~D.\ 2005, \nat, 435, 317 

\bibitem[Mamajek(2005)]{2005ApJ...634.1385M} Mamajek, E.~E.\ 2005, \apj, 
634, 1385

\bibitem[Mandell et al.(2008)]{2008ApJ...681L..25M} Mandell, A.~M., Mumma, 
M.~J., Blake, G.~A., Bonev, B.~P., Villanueva, G.~L., 
\& Salyk, C.\ 2008, \apjl, 681, L25 

\bibitem[Marsh 
\& Mahoney(1992)]{1992ApJ...395L.115M} Marsh, K.~A., \& Mahoney, M.~J.\ 1992, \apjl, 395, L115

\bibitem[Muzerolle et al.(2000)]{2000ApJ...535L..47M} Muzerolle, J., 
Calvet, N., Brice{\~n}o, C., Hartmann, L., 
\& Hillenbrand, L.\ 2000, \apjl, 535, L47 

\bibitem[Muzerolle et al.(2010)]{2010ApJ...708.1107M} Muzerolle, J., Allen, 
L.~E., Megeath, S.~T., Hern{\'a}ndez, J., 
\& Gutermuth, R.~A.\ 2010, \apj, 708, 1107 

\bibitem[Najita et al.(1996)]{1996ApJ...462..919N} Najita, J., Carr, J.~S., 
Glassgold, A.~E., Shu, F.~H., \& Tokunaga, A.~T.\ 1996, \apj, 462, 919

\bibitem[Najita et al.(2007)]{2007MNRAS.378..369N} Najita, J.~R., Strom, 
S.~E., \& Muzerolle, J.\ 2007, \mnras, 378, 369 

\bibitem[Najita et al.(2010)]{2010ApJ...712..274N} Najita, J.~R., Carr, 
J.~S., Strom, S.~E., Watson, D.~M., Pascucci, I., Hollenbach, D., Gorti, 
U., \& Keller, L.\ 2010, \apj, 712, 274 

\bibitem[Owen et al.(2010)]{2010MNRAS.401.1415O} Owen, J.~E., Ercolano, B., 
Clarke, C.~J., \& Alexander, R.~D.\ 2010, \mnras, 401, 1415 

\bibitem[Pascucci 
\& Sterzik(2009)]{2009ApJ...702..724P} Pascucci, I., \& Sterzik, M.\ 2009, \apj, 702, 724 

\bibitem[Pascucci et al.(2011)]{2011ApJ...XYX..XYZN} Pascucci, I., et al.\ 2011, \apj, submitted 

\bibitem[Pontoppidan et al.(2008)]{2008ApJ...684.1323P} Pontoppidan, K.~M., 
Blake, G.~A., van Dishoeck, E.~F., Smette, A., Ireland, M.~J., 
\& Brown, J.\ 2008, \apj, 684, 1323 

\bibitem[Pontoppidan et al.(2010)]{2010ApJ...720..887P} Pontoppidan, K.~M., 
Salyk, C., Blake, G.~A., Meijerink, R., Carr, J.~S., 
\& Najita, J.\ 2010, \apj, 720, 887 

\bibitem[Pringle(1981)]{1981ARA&A..19..137P} Pringle, J.~E.\ 1981, \araa, 19, 137

\bibitem[Qi et al.(2004)]{2004ApJ...616L..11Q} Qi, C., et al.\ 2004, \apjl, 
616, L11 

\bibitem[Qi et al.(2008)]{2008ApJ...681.1396Q} Qi, C., Wilner, D.~J., 
Aikawa, Y., Blake, G.~A., \& Hogerheijde, M.~R.\ 2008, \apj, 681, 1396 

\bibitem[Qi et al.(2006)]{2006ApJ...636L.157Q} Qi, C., Wilner, D.~J., 
Calvet, N., Bourke, T.~L., Blake, G.~A., Hogerheijde, M.~R., Ho, P.~T.~P., 
\& Bergin, E.\ 2006, \apjl, 636, L157

\bibitem[Rafikov 
\& De Colle(2006)]{2006ApJ...646..275R} Rafikov, R.~R., \& De Colle, F.\ 2006, \apj, 646, 275 

\bibitem[Ratzka et 
al.(2007)]{2007A&A...471..173R} Ratzka, T., Leinert, C., Henning, T., Bouwman, J., Dullemond, C.~P., \& Jaffe, W.\ 2007, \aap, 471, 173 

\bibitem[Rettig et al.(2004)]{2004ApJ...616L.163R} Rettig, T.~W., Haywood, 
J., Simon, T., Brittain, S.~D., \& Gibb, E.\ 2004, \apjl, 616, L163 

\bibitem[Rice et al.(2003)]{2003MNRAS.342...79R} Rice, W.~K.~M., Wood, K., 
Armitage, P.~J., Whitney, B.~A., \& Bjorkman, J.~E.\ 2003, \mnras, 342, 79 

\bibitem[Richter et al.(2003)]{2003SPIE.4857...37R} Richter, M.~J., Lacy, 
J.~H., Zhu, Q., Jaffe, D.~T., Greathouse, T.~K., Moerchen, M., Mar, D.~J., 
\& Knez, C.\ 2003, \procspie, 4857, 37

\bibitem[Robrade 
\& Schmitt(2006)]{2006A&A...449..737R} Robrade, J., \& Schmitt, J.~H.~M.~M.\ 2006, \aap, 449, 737 

\bibitem[Salyk et al.(2007)]{2007ApJ...655L.105S} Salyk, C., Blake, G.~A., 
Boogert, A.~C.~A., \& Brown, J.~M.\ 2007, \apjl, 655, L105

\bibitem[Salyk et al.(2008)]{2008ApJ...676L..49S} Salyk, C., Pontoppidan, 
K.~M., Blake, G.~A., Lahuis, F., van Dishoeck, E.~F., 
\& Evans, N.~J., II 2008, \apjl, 676, L49

\bibitem[Scoville et al.(1980)]{1980ApJ...240..929S} Scoville, N.~Z., 
Krotkov, R., \& Wang, D.\ 1980, \apj, 240, 929 

\bibitem[Setiawan et al.(2008)]{2008Natur.451...38S} Setiawan, J., Henning, 
T., Launhardt, R., M{\"u}ller, A., Weise, P., K{\"u}rster, M.\ 2008, \nat, 451, 38 

\bibitem[Sicilia-Aguilar et al.(2010)]{2010ApJ...710..597S} 
Sicilia-Aguilar, A., Henning, T., \& Hartmann, L.~W.\ 2010, \apj, 710, 597 

\bibitem[Skrutskie et al.(1990)]{1990AJ.....99.1187S} Skrutskie, M.~F., 
Dutkevitch, D., Strom, S.~E., Edwards, S., Strom, K.~M., 
\& Shure, M.~A.\ 1990, \aj, 99, 1187

\bibitem[Stelzer 
\& Schmitt(2004)]{2004A&A...418..687S} Stelzer, B., \& Schmitt, J.~H.~M.~M.\ 2004, \aap, 418, 687

\bibitem[Storzer 
\& Hollenbach(1998)]{1998ApJ...502L..71S} Storzer, H., \& Hollenbach, D.\ 1998, \apjl, 502, L71 

\bibitem[Strom et al.(1989)]{1989AJ.....97.1451S} Strom, K.~M., Strom, 
S.~E., Edwards, S., Cabrit, S., \& Skrutskie, M.~F.\ 1989, \aj, 97, 1451

\bibitem[Takeuchi et al.(1996)]{1996ApJ...460..832T} Takeuchi, T., Miyama, 
S.~M., \& Lin, D.~N.~C.\ 1996, \apj, 460, 832 

\bibitem[Tappe et al.(2008)]{2008ApJ...680L.117T} Tappe, A., Lada, C.~J., 
Black, J.~H., \& Muench, A.~A.\ 2008, \apjl, 680, L117 

\bibitem[Thi et 
al.(2004)]{2004A&A...425..955T} Thi, W.-F., van Zadelhoff, G.-J., \& van Dishoeck, E.~F.\ 2004, \aap, 425, 955

\bibitem[Thi et 
al.(2010)]{2010A&A...518L.125T} Thi, W.-F., et al.\ 2010, \aap, 518, L125 
\bibitem[Tsuji(1966)]{1966PASJ...18..127T} Tsuji, T.\ 1966, \pasj, 18, 127

\bibitem[Turner et al.(2010)]{2010ApJ...708..188T} Turner, N.~J., 
Carballido, A., \& Sano, T.\ 2010, \apj, 708, 188

\bibitem[Valenti et al.(2003)]{2003ApJS..147..305V} Valenti, J.~A., Fallon, 
A.~A., \& Johns-Krull, C.~M.\ 2003, \apjs, 147, 305 

\bibitem[van Dishoeck 
\& Dalgarno(1984)]{1984Icar...59..305V} van Dishoeck, E.~F., \& Dalgarno, A.\ 1984, \icarus, 59, 305 

\bibitem[van Dishoeck 
\& Dalgarno(1984)]{1984ApJ...277..576V} van Dishoeck, E.~F., \& Dalgarno, A.\ 1984, \apj, 277, 576

\bibitem[van Dishoeck et 
al.(2003)]{2003A&A...400L...1V} van Dishoeck, E.~F., Thi, W.-F., \& van Zadelhoff, G.-J.\ 2003, \aap, 400, L1

\bibitem[van Harrevelt 
\& van Hemert(2000)]{2000JChPh.112.5777V} van Harrevelt, R., \& van Hemert, M.~C.\ 2000, \jcp, 112, 5777 

\bibitem[van Zadelhoff et al.(2001)]{2001A&A...377..566V} van Zadelhoff, G.-J., van Dishoeck, E.~F., Thi, W.-F., \& Blake, G.~A.\ 2001, \aap, 377, 566 

\bibitem[Varni{\`e}re et al.(2006)]{2006ApJ...640.1110V} Varni{\`e}re, P., 
Blackman, E.~G., Frank, A., \& Quillen, A.~C.\ 2006, \apj, 640, 1110

\bibitem[Webb et al.(1999)]{1999ApJ...512L..63W} Webb, R.~A., Zuckerman, 
B., Platais, I., Patience, J., White, R.~J., Schwartz, M.~J., 
\& McCarthy, C.\ 1999, \apjl, 512, L63 

\bibitem[Weintraub et al.(2000)]{2000ApJ...541..767W} Weintraub, D.~A., 
Kastner, J.~H., \& Bary, J.~S.\ 2000, \apj, 541, 767 

\bibitem[Wilner et al.(2003)]{2003ApJ...596..597W} Wilner, D.~J., Bourke, 
T.~L., Wright, C.~M., J{\o}rgensen, J.~K., van Dishoeck, E.~F., 
\& Wong, T.\ 2003, \apj, 596, 597

\bibitem[Zuckerman et al.(1995)]{1995Natur.373..494Z} Zuckerman, B., 
Forveille, T., \& Kastner, J.~H.\ 1995, \nat, 373, 494

\begin{table}
\caption{Observations of gas emission lines}
\label{obsdata}
\begin{tabular}{lrcll}
\hline
Line & $\lambda$ & Line Luminosity (l$_{\odot}$) & Reference & Obs. Facility \\
\hline
CO 6-5 & 434$\mu$m  & $4.6\times 10^{-8}$ & Qi et al. 2006 & SMA \\
CO 3-2 & 867$\mu$m & $2.4\times 10^{-8}$ & Qi et al. 2006 & SMA \\
CO 2-1 & 1.3 mm &$7.9\times 10^{-9}$ & Qi et al. 2006 & SMA \\
HCO$^+$4-3 & 841$\mu$m & $7.7 \times10^{-9}$ & van Zadelhoff et al. 2001 & JCMT\\ 
OI & 63$\mu$m & $2.7\times 10^{-6}$ & Thi et al. 2010 & Herschel PACS \\
OI & 145$\mu$m & $ < 4\times 10^{-7}$ & Thi et al. 2010 & Herschel PACS \\
CII & 157$\mu$m & $<4.8\times 10^{-7}$ & Thi et al. 2010 & Herschel PACS \\
\hline
NeII &  12.8 $\mu$m &  $3.9-4.7 \times 10^{-6}$ & Najita et al. 2010	& Spitzer IRS \\
       &   & $3.0\times 10^{-6}$ & Pascucci \& Sterzik 2009 & VLT/VISIR ($r<10$AU)\\
NeIII & 15.5 $\mu$m &  $2.0\times 10^{-7}$  & Najita et al. 2010	& Spitzer IRS \\ 
H$_2$ S(2) & 12.28 $\mu$m & $4.8 \times 10^{-7}$ & Najita et al. 2010	& Spitzer IRS \\
                   &                         & $<4.8 \times 10^{-7}$ & Bitner et al. 2008 & Gemini/TEXES ($r<13.7$AU) \\
H$_2$ S(1) & 17.0 $\mu$m & $0.96-1.0 \times 10^{-6}$ & Najita et al. 2010	& Spitzer IRS \\
                   &                        & $<5.5\times 10^{-7}$ & Bitner et al. 2008 & Gemini/TEXES ($r<20$AU) \\
OH & MIR- SH & $6.2\times 10^{-6} $  & Najita et al. 2010	& Spitzer IRS \\ 
 \hline
 CO$_{vib}$ & NIR &$1.0\times 10^{-5}$ & Salyk et al. 2007 & Keck II/Echelle\\
 H$_{2,vib}$ & 2.12$\mu$m & $7.9\times 10^{-8}$ & Bary et al. 2003 & KPNO/Phoenix \\
 H$_2$ fluo. & UV & $1.9\times 10^{-4}$ & Herczeg et al. 2002 & HST/STIS  \\
  OI & 6300 A & $1.0 \times 10^{-5} $ &Suzan Edwards & Keck/Echelle \\
 OI & 5577 A & $ 1.4 \times 10^{-6}$  &Suzan Edwards & Keck/Echelle\\
 SII & 4069 A & $ 3.0\times10^{-6}$ & Pascucci et al. 2011 & FEROS \\
\hline
\hline
\end{tabular}
\end{table}

\begin{table}
\caption{Dust model parameters}
\begin{tabular}{ll}
\hline
{\sc Optically Thin Inner Disk} &\\
\hline
Inner radius & 0.06 AU \\
Outer radius & 3.5 AU \\
Dust disk mass  & $2.4\times10^{25}$g\\
Surface Density $\Sigma$ & $\propto r^{-1}$ \\
Min. grain size & $0.9\mu$m\\
Max. grain size & $2.0\mu$m \\
Dust Composition & Glassy Pyroxene \\
\hline
{\sc Optically Thick Outer Disk} \\
\hline
Inner radius & 3.5 AU \\
Outer radius & 200 AU \\
Dust disk mass (with $a<1$mm) & $4.8\times10^{29}$g \\
Surface Density $\Sigma$ & Composite (Fig.1)  \\
Min. grain size ($a_{min}$) & $0.01\mu$m \\
Max. grain size ($a_{max}$) & $1$ mm \\
Dust Composition & Silicates$+$graphite \\
\hline
\end{tabular}
\label{calvetdust}
\end{table}

\begin{table}
\label{idtable}
\caption{Model calculations for lines (sum of inner and outer disk contributions) that constrain the inner disk mass}
\begin{tabular}{l | c | c | c | c | c}
\hline
\hline
Line & Obs.Data  &SMH & SML & & Contribution\\
& &  $(M_i=10^{-4}$\ms)& $ (M_i=10^{-5}$\ms) & ($M_i=10^{-6}$\ms) & from $r>4$AU \\
&(L$_{\odot}$) & (L$_{\odot}$) &(L$_{\odot}$)&(L$_{\odot}$)&(L$_{\odot}$) \\
\hline
CO$_{vib}$ (P Series) & $10^{-5}$ & $2.1\times10^{-5}$ & $1.8\times10^{-5}$ &$1.6\times10^{-5}$ &   --- \\
OH (prompt) & $1.5\times10^{-5}$ & $  7.8\times 10^{-6 } $ & $  4\times 10^{-6 } $  &$2.8\times10^{-6}$ &$ 2.8 \times 10^{- 6} $\\
 & & & &&\\
OI 6300\AA  & $  10^{-5 } $ & $1.5\times10^{-5}$ & $8.4\times10^{-6}$ & $8.9\times10^{-6}$& $6.1\times 10^{-6}$ \\   
{\it \ (Thermal})   &&$  (3.1\times 10^{-6 }) $&$  (2.3\times 10^{-6 }) $&$(2.8\times10^{-6}) $ &$ (1.1 \times 10^{-6 }) $\\
{\it \ (Prompt})   && $  (1.2\times 10^{-5 }) $ & $  (6.1\times 10^{-6 }) $ &$(6.1\times10^{-6})$& $  (5\times 10^{- 6}) $ \\
& & & &&\\
OI 5577\AA  & $  1.4\times 10^{-6 } $ &$1.92\times10^{-6}$ & $1.6\times10^{-6}$&$9.4\times10^{-7} $ & $7\times 10^{-7}$\\
{\it \ (Thermal}) & & $  (2.2\times 10^{-7 }) $&$  (1.0\times 10^{-7 }) $& $(1.3\times10^{-7}) $&$  (1.0\times 10^{-10 }) $ \\
{\it \ (Prompt})  & & $  (1.7\times 10^{-6}) $ & $  (1.5\times 10^{-6 }) $ &$(8.1\times10^{-7})$& $ (7 \times 10^{-7 }) $\\
& & & &&\\
H$_2$ S(2) &$  4.8\times 10^{-7 } $&$  6.8\times 10^{-7 } $&$  3.4\times 10^{-7 } $& $2.5\times10^{-7}$ &$  2.2\times 10^{-7 } $ \\
H$_2$ S(1) &$  1.0\times 10^{-6 } $&$  1.3\times 10^{-6 } $&$  1.2\times 10^{-6 } $& $1.1\times10^{-6}$&$  1.1\times 10^{-6 } $\\
NeII &$  5.0\times 10^{-6 } $&$  3.1\times 10^{-6 } $&$  3.0\times 10^{-6 } $& $3.4\times10^{-6}$&$  2.3\times 10^{-6 } $\\
NeIII &$  2.0\times 10^{-7 } $&$  3.0\times 10^{-7 } $&$  3.0\times 10^{-7 } $&$3.5\times10^{-7} $&$  1.0\times 10^{-7 } $\\
\hline
OI linewidth (km s$^{-1}$) & 10.0 &  11.0  & 6.0&6.1 &\\
$\alpha$ (visc. parameter) & & 0.01-0.02 & 0.1-0.2 & 1-2&  \\
\hline
\end{tabular}
\end{table}

\newpage
\begin{table}
\label{SMLtable}
\caption{Calculated model line luminosities for lines originating beyond 4AU}
\begin{tabular}{lcc}
\hline
Line ($\lambda$) & Obs. Data & Model \\ 
& (L$_{\odot}$) & (L$_{\odot}$) \\
\hline 
OI(63$\mu$m)  & $2.7\times10^{-6}$& $3-6\times10^{-6}$ \\
OI(145$\mu$m)  & $ <4\times10^{-7}$& $1-3\times10^{-7}$ \\
CII(158$\mu$m)  & $<4.5\times10^{-6}$& $3\times10^{-6}$ \\
CO 6-5(430$\mu$m) & $4.6\times 10^{-8}$ & $8 \times 10^{-8}$  \\
CO 3-2(866$\mu$m) & $2.4\times 10^{-8}$ & $ 2.4 \times 10^{-8}$ \\
CO 2-1(1.3mm) &$7.9\times 10^{-9}$ &  $ 7.1 \times 10^{-9}$ \\
HCO$^+4-3$(841$\mu$m) & $7.7\times10^{-9}$ & $4\times10^{-9}$ \\
\hline
\end{tabular}
\end{table}

\begin{figure}
\plotone{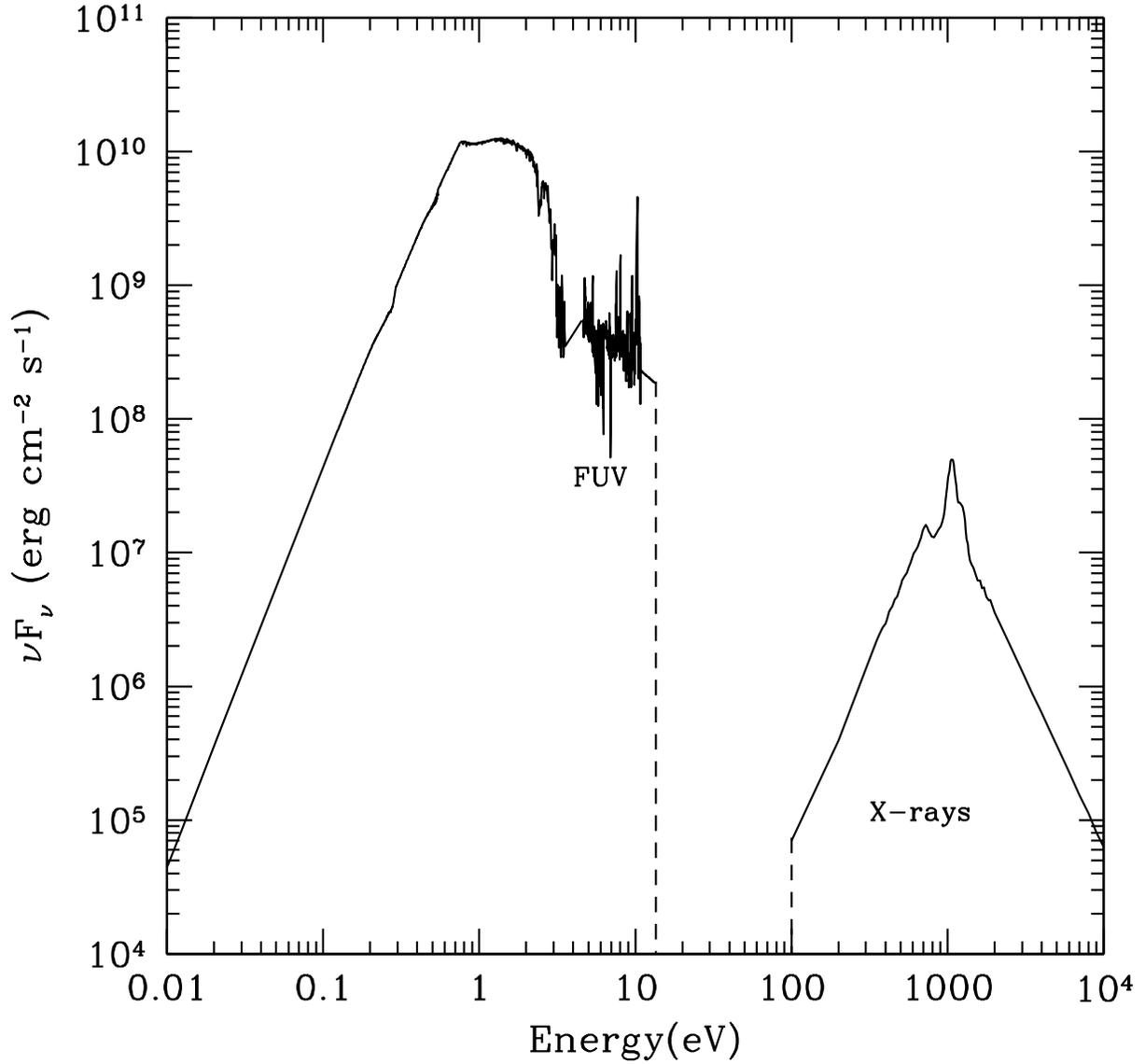}
\caption{The input spectrum used for TW Hya, which is a composite of the stellar, FUV and X-ray spectra shown here as the flux at the stellar surface as a function of photon energy.  The vertical dashed lines indicate the EUV region which has been ignored in our modeling. }
\label{spec}
\end{figure}

\begin{figure}
\plotone{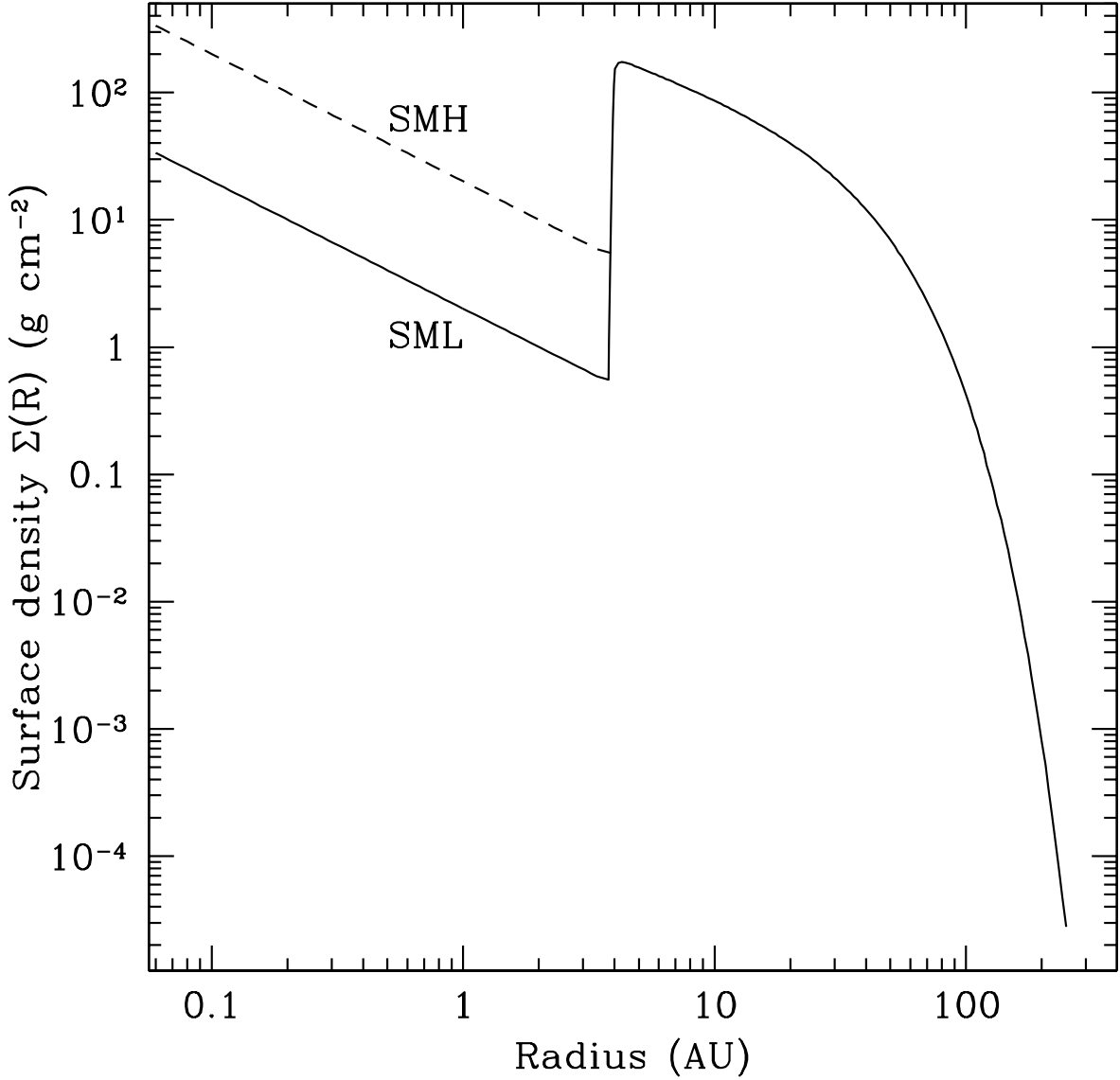}
\caption{Surface density distribution of the SML (solid line) and SMH (dashed line) disks. The surface density follows a power law ($\Sigma \propto r^{-1}$) within 4 AU, rises over a ``rim" region ($3.5-4$AU), and is approximately a power law ($\Sigma \propto r^{-0.7}$) from $4-40$AU. In the outer disk ($r\gtrsim40$AU), the surface density rapidly decreases with radius. See \S 3.1 for the analytic formulation of $\Sigma(r)$.}
\label{sigmar}
\end{figure}

\begin{figure}
\plottwo{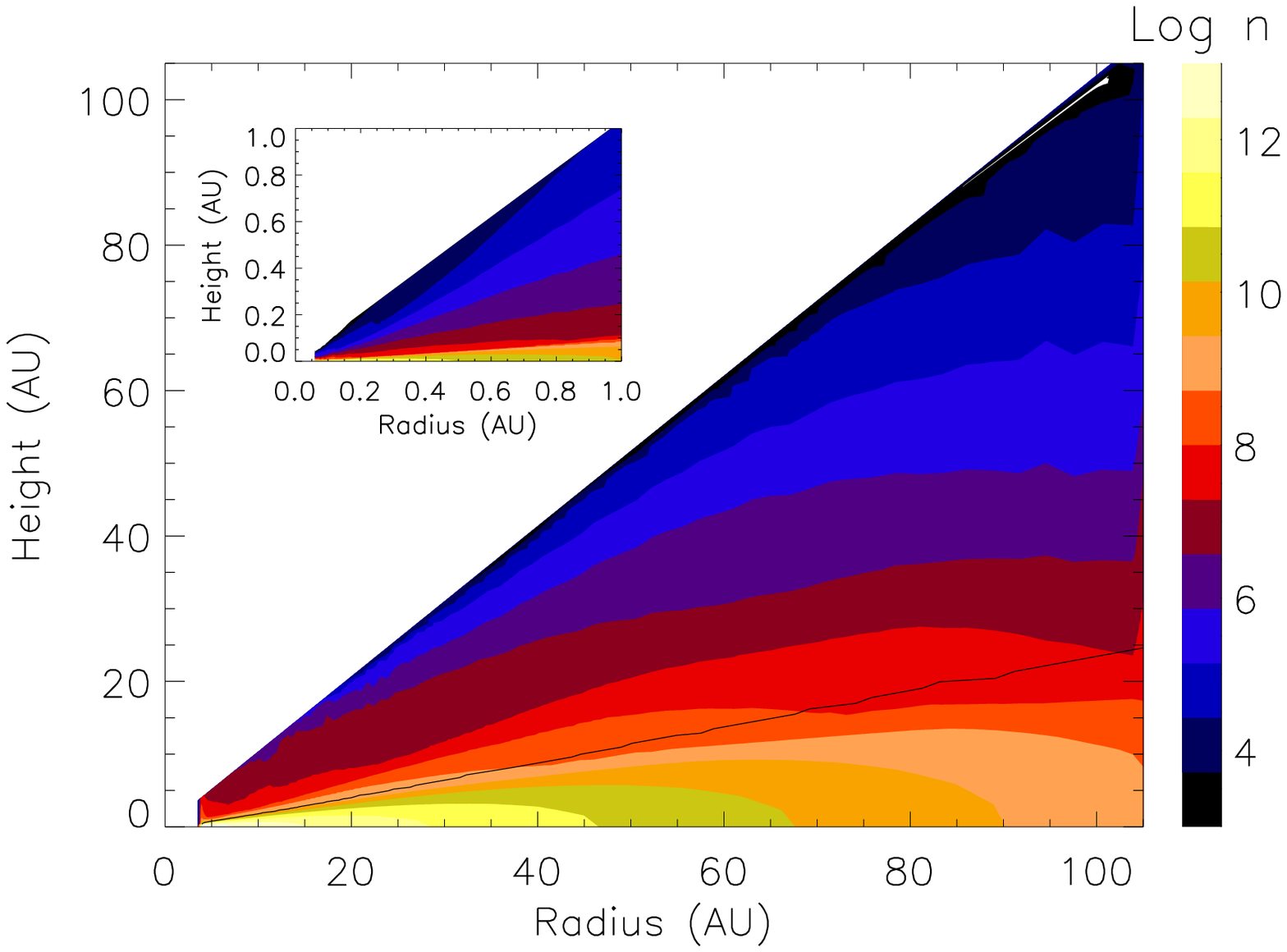}{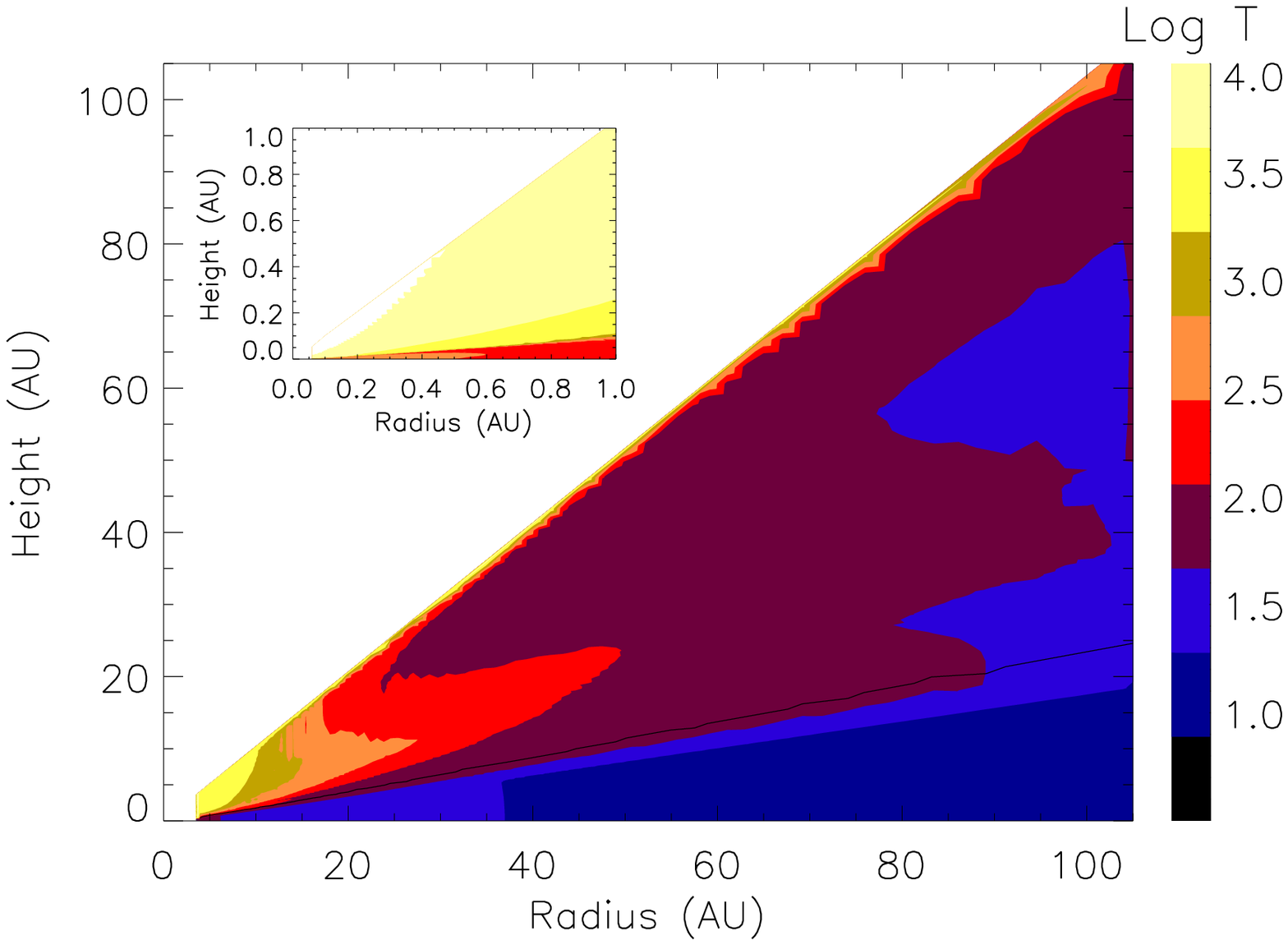}
\caption{The density and temperature structure of the disk for the SML disk. The outer disk ($r\gtrsim 40$AU) density structure shows a somewhat flattened disk structure, resulting in relatively low disk temperatures in these regions. The black line in both panels shows the height at which the visual extinction to the star, A$_{\rm V}=1$. The insets show the structure in the inner disk for $r<1$AU where most of the line emission is produced. }
\label{nt-contour}
\end{figure}

\begin{figure}
\plotone{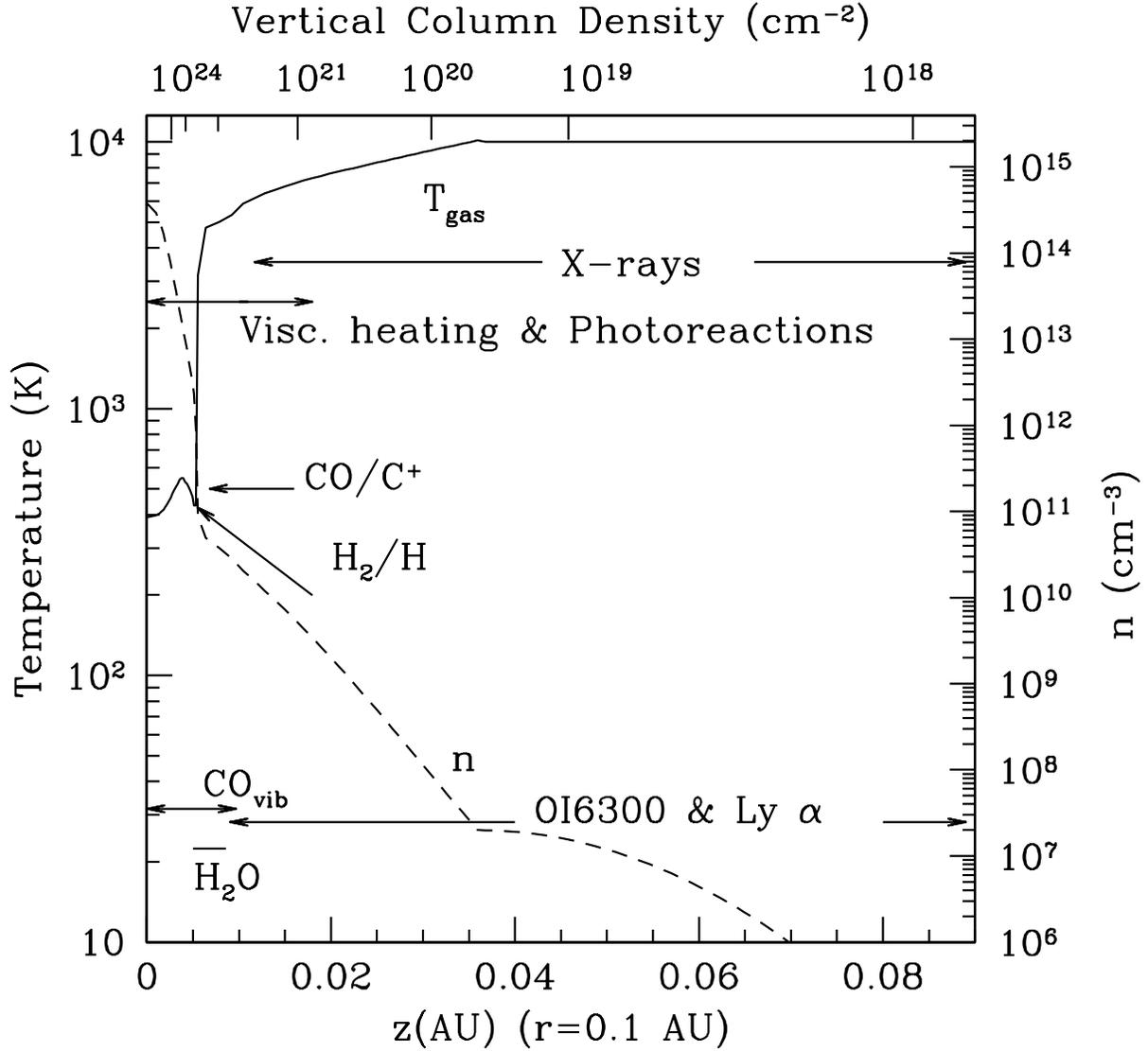}
\caption{ The vertical temperature(solid line) and density(dashed line) distribution is shown at a typical radius in the inner disk (0.1AU). The dominant heating (above) and cooling (below) mechanisms are shown and the radial extent where they are important in overall thermal balance is indicated. The  photodissociation fronts of CO and H$_2$ are also marked on the plots.
}
\label{vslices1}
\end{figure}

\begin{figure}
\plotone{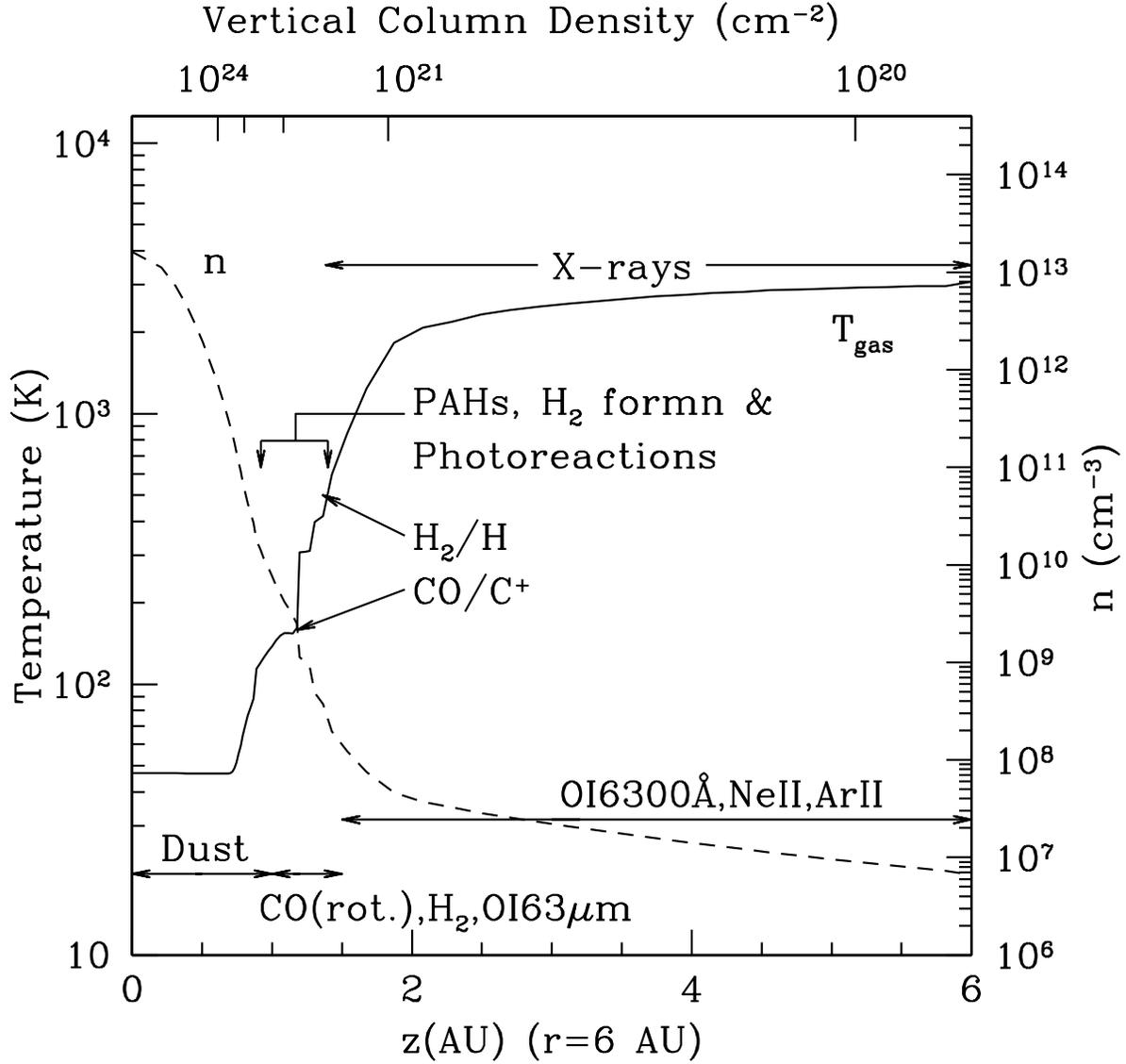}
\caption{The disk structure in the mid-disk region at a radius of 6 AU. Here, the dominant FUV heating by PAHs and photoreactions near the A$_V=1$ layer, and X-rays at the surface. Near the midplane, dust collisions are important. The main gas coolants are CO and H$_2$ rotational lines deeper in the disk, and OI (63$\mu$m and 6300\AA) and [NeII] and [ArII] at the disk surface.}
\label{vslices2}
\end{figure}

\begin{figure}
\plotone{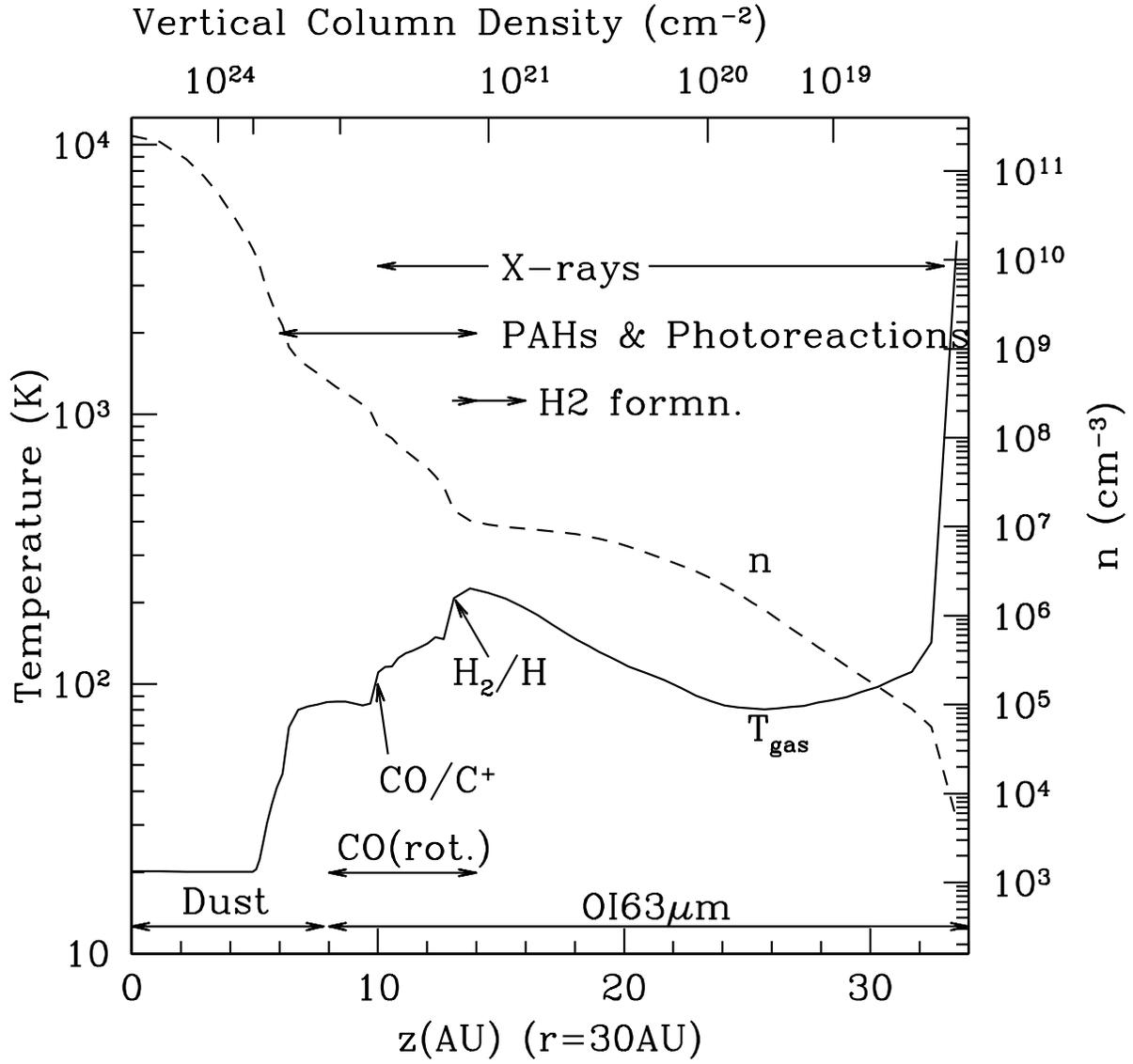}
\caption{The disk structure at a representative radius in the outer disk (30AU). The dominant heating and cooling mechanisms and their radial extent is indicated in the figure. }
\label{vslices3}
\end{figure}

\begin{figure}
\plotone{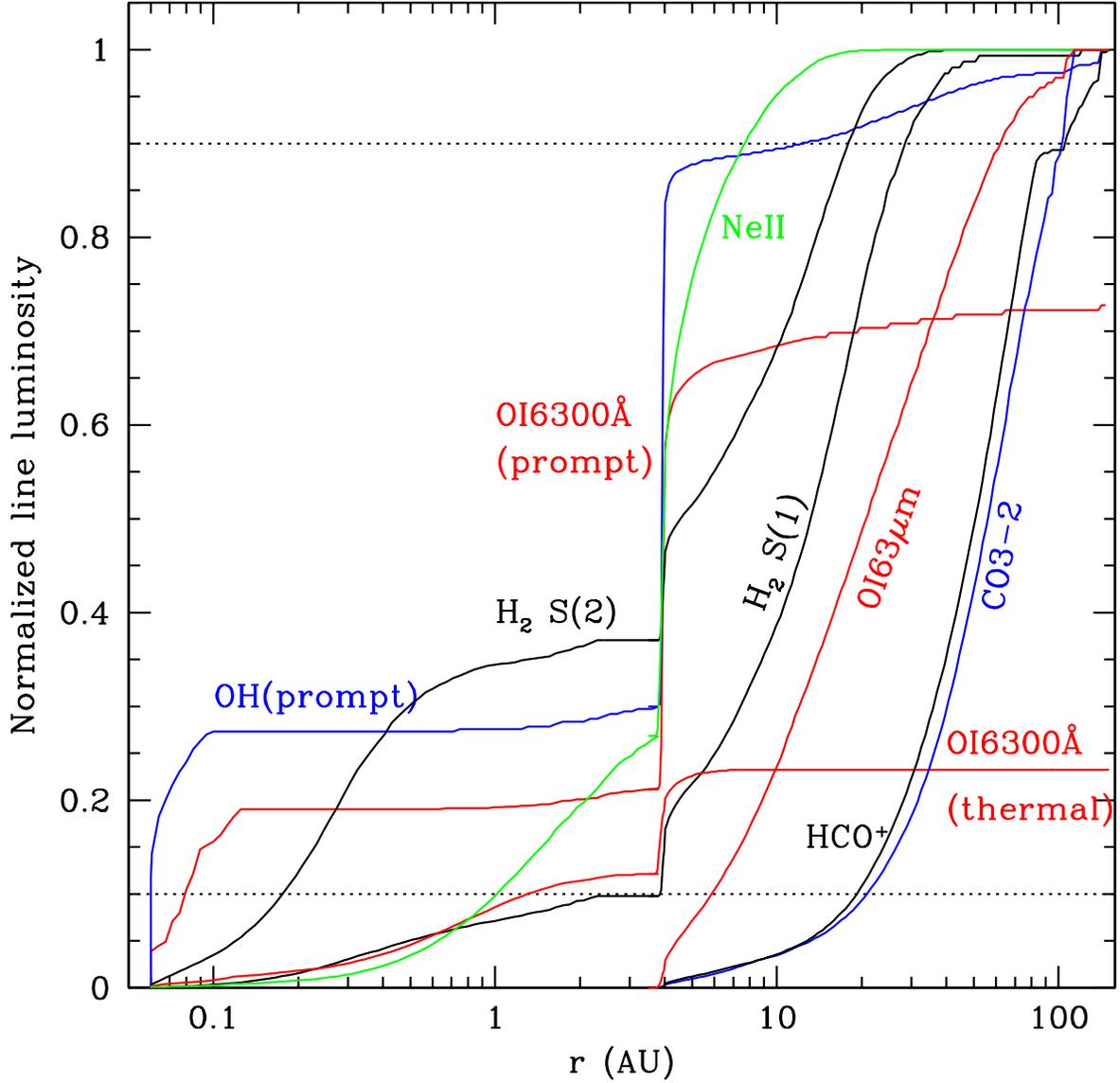}
\caption{The cumulative line luminosities ($L(r)$) normalized to their total values (Table~\ref{idtable}) for the SML as a function of disk radius. The molecular lines, including prompt emission, usually arise from near the A$_V=1$ layer, whereas the OI and NeII thermal emission originate above this layer in the disk surface.  The 10\% and 90\% levels (dotted lines) mark the radii between which most of the emission originates. For example, the [NeII]12.8$\mu$m line mainly arises from gas located between 1 and 8 AU.}
\label{lines}
\end{figure}

\begin{figure}
\plotone{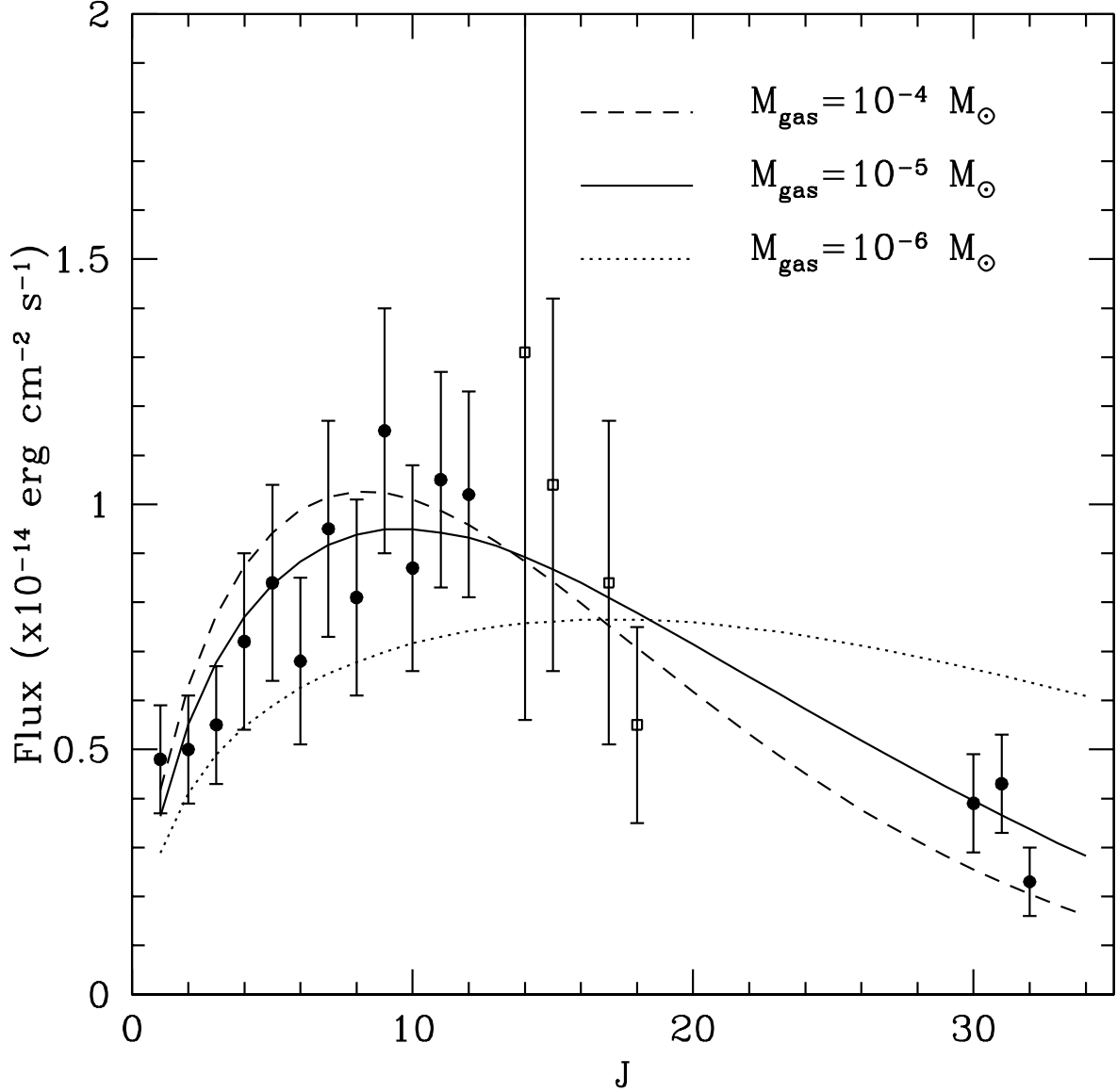}
\caption{Calculated CO rovibrational fluxes for the $v=1-0,$ P(J$=1-30$) transitions and the SML inner disk mass (solid line) are shown for each rotational quantum number J. The filled circles are the photosphere and veiling corrected measurements from Salyk et al. (2007). Open squares are corrected data from Rettig et al. (2004) (see text). The dashed line is for a 
  higher disk mass ($10^{-4}$\ms) that also fits the data reasonably well. The $10^{-6}$\ms\ inner disk mass model (dotted line)  is a poor fit to the data.}
\label{covib}
\end{figure}

\begin{figure}
\plotone{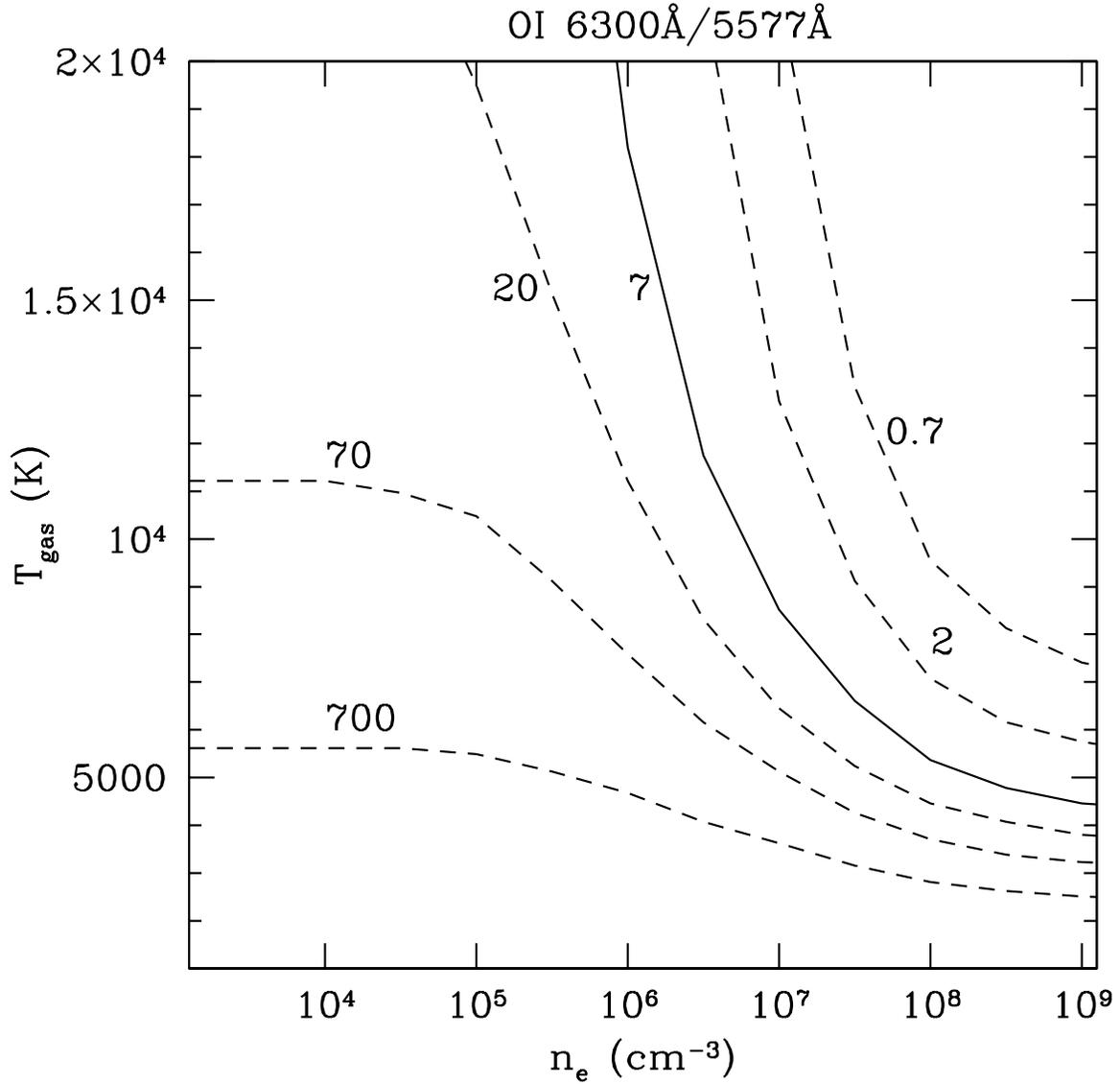}
\caption{Contours of the  thermal line ratio OI6300\AA/OI5577\AA\  as a function of the gas temperature and electron density from a statistical equilibrium calculation of the O atomic levels are shown. TW Hya has an observed ratio of 7 (solid line) which requires high electron densities at typical neutral gas temperatures. For an electron abundance $x_e\sim 0.01$ and $T\sim 8000$K, gas densities $n\sim 10^{9}$cm$^{-3}$ are needed to reproduce the observed 6300\AA/5577\AA\ line ratio by thermal OI emission. }
\label{o1ratio}
\end{figure}

\end{document}